\documentclass[twocolumn]{aastex63}

\newcounter{daggerfootnote}

\usepackage{amsmath}
\usepackage{color}

\newcommand{\smm}{\sum\limits}

\newcommand{\mebv}{E(B-V)}

\newcommand{\mgmag}{m}
\newcommand{\gmag}{$\mgmag$}
\newcommand{\mragal}{\alpha_i} 
\newcommand{\ragal}{$\mragal$}
\newcommand{\mdecgal}{\delta_i} 
\newcommand{\decgal}{$\mdecgal$}
\newcommand{\mgalhalflight}{\phi_{i}} 
\newcommand{\galhalflight}{$\mgalhalflight$}

\newcommand{\photutils}{{\sc photutils}}

\newcommand{\mNc}{N_c} 
\newcommand{\Nc}{$\mNc$}

\newcommand{\mrafrb}{\alpha_{\rm FRB}} 
\newcommand{\rafrb}{$\mrafrb$}
\newcommand{\mdecfrb}{\delta_{\rm FRB}} 
\newcommand{\decfrb}{$\mdecfrb$}
\newcommand{\meefrb}{\epsilon_{\rm FRB}} 
\newcommand{\eefrb}{$\meefrb$}
\newcommand{\msigmafrb}{\sigma_{\rm FRB}} 
\newcommand{\sigmafrb}{$\msigmafrb$}
\newcommand{\meea}{\epsilon_a}  
\newcommand{\eea}{$\meea$}
\newcommand{\meeb}{\epsilon_b}  
\newcommand{\eeb}{$\meeb$}
\newcommand{\meep}{\epsilon_{\rm PA}}  
\newcommand{\eep}{$\meep$}
\newcommand{\nsecure}{nine}  
\newcommand{\frblocal}{99.9\%}  

\newcommand{\msep}{\theta} 
\newcommand{\sep}{$\msep$}
\newcommand{\mhalflight}{\phi} 
\newcommand{\halflight}{$\mhalflight$}
\newcommand{\aimage}{\texttt{a\_image}}

\newcommand{\mtheff}{\theta_{\rm eff}} 
\newcommand{\theff}{$\mtheff$}
\newcommand{\mpchance}{P^c}
\newcommand{\pchance}{$\mpchance$}

\newcommand{\mnumden}{\Sigma(\mgmag)}
\newcommand{\numden}{$\mnumden$}
\newcommand{\minumden}{\Sigma(\mgmag_i)}

\newcommand{\mthmax}{\theta_{\rm max}}  
\newcommand{\thmax}{$\mthmax$}
\newcommand{\mpoffset}{p(\omega|O_i)}  
\newcommand{\poffset}{$\mpoffset$}
\newcommand{\mPO}{P(O)}  
\newcommand{\PO}{$\mPO$}
\newcommand{\mPOi}{P(O_i)}  
\newcommand{\POi}{$\mPOi$}
\newcommand{\mPU}{P(U)}  
\newcommand{\PU}{$\mPU$}
\newcommand{\uguess}{0}  
\newcommand{\nsets}{two}  

\newcommand{\mpxOi}{p(x|O_i)}  

\newcommand{\adopted}{adopted}

\newcommand{\mPOx}{P(O|x)}  
\newcommand{\POx}{$\mPOx$}
\newcommand{\mPOix}{P(O_i|x)}  
\newcommand{\POix}{$\mPOix$}
\newcommand{\mPUx}{P(U|x)}  
\newcommand{\PUx}{$\mPUx$}
\newcommand{\mPOsec}{P_{\rm secure}}  

\newcommand{\POvsec}{0.95}  

\newcommand{\simsecure}{35}  

\newcommand{\mdmcosmic}{{\rm DM}_{\rm cosmic}}
\newcommand{\dmcosmic}{$\mdmcosmic$}

\newcommand{\mdmfrb}{{\rm DM}_{\rm FRB}}

\newcommand{\mdmhost}{{\rm DM}_{\rm host}}
\newcommand{\dmhost}{$\mdmhost$}

\newcommand{\mdmmwism}{{\rm DM}_{\rm MW,ISM}}
\newcommand{\dmmwism}{$\mdmmwism$}

\begin{document}

\title{Probabilistic Association of Transients to their Hosts (PATH)}

\correspondingauthor{J. Xavier Prochaska}
\email{xavier@ucolick.org}

\author[0000-0002-2059-0525]{Kshitij Aggarwal}
\affil{Department of Physics and Astronomy, West Virginia University, Morgantown, WV 26506, USA}
\affil{Center for Gravitational Waves and Cosmology, West Virginia University, Chestnut Ridge Research Building, Morgantown, WV, USA}

\author[0000-0002-7034-4621]{Tam\'as Budav\'ari}
\affil{Department of Applied Mathematics and Statistics, Johns Hopkins University, Baltimore, MD, USA}
\affil{Department of Physics and Astronomy, Johns Hopkins University, Baltimore, MD, USA}

\author[0000-0001-9434-3837]{Adam T. Deller}
\affil{Centre for Astrophysics and Supercomputing, Swinburne University of Technology, Hawthorn, VIC 3122, Australia}

\author[0000-0003-0307-9984]{Tarraneh Eftekhari}
\affil{Center for Astrophysics $|$ Harvard \& Smithsonian, Cambridge, MA 02138, USA}

\author[0000-0002-6437-6176]{Clancy W. James}
\affil{International Centre for Radio Astronomy Research, Curtin University, Bentley, WA 6102, Australia}

\author[0000-0002-7738-6875]{J. Xavier Prochaska}
\affil{University of California - Santa Cruz, 1156 High St., Santa Cruz, CA, USA 95064}
\affil{Kavli Institute for the Physics and Mathematics of the Universe (Kavli IPMU), 5-1-5 Kashiwanoha, Kashiwa, 277-8583, Japan}

\author[0000-0003-2548-2926]{Shriharsh P. Tendulkar}
\affiliation{Department of Astronomy and Astrophysics, Tata Institute of Fundamental Research, Homi Bhabha Road, Colaba, Mumbai, Maharashtra, 400005, India}
\affiliation{National Centre for Radio Astrophysics, Pune University Campus, Post Bag 3, Ganeshkhind, Pune, Maharashtra, 411007, India}

\shorttitle{FRB host galaxy associations}
\shortauthors{FRB Association Team}

\received{\today}
\revised{\today}
\accepted{\today}
\submitjournal{ApJ}

\begin{abstract}
We introduce a new method to estimate the probability
that an extragalactic transient source 
is associated with a candidate host galaxy.  This approach relies
solely on simple observables: 
sky coordinates and their uncertainties, 
galaxy fluxes and angular sizes.
The formalism invokes Bayes' rule to calculate the
posterior probability \POix\ from the galaxy prior
\PO, observables $x$, and an assumed 
model for the true distribution
of transients in/around their host galaxies.
Using simulated transients placed in the well-studied 
COSMOS field, we consider several agnostic and 
physically motivated priors and offset distributions
to explore the method sensitivity.
We then apply the methodology to the set of 
13~fast radio bursts (FRBs) localized with an uncertainty of several
arcseconds.  Our methodology finds 
\nsecure\ of these are securely associated to a
single host galaxy, $\mPOix>0.95$.
We examine the observed and intrinsic properties of
these secure FRB hosts, 
recovering similar  distributions as previous works.
Furthermore, we find a strong correlation between 
the apparent magnitude of the securely identified 
host galaxies and the estimated cosmic dispersion 
measures of the corresponding FRBs, which results
from the Macquart relation.
Future work with FRBs will leverage this relation and
other measures from the secure hosts as priors for
future associations.
The methodology is generic to transient type,
localization error, and image quality.
We encourage its application to other transients
where host galaxy associations are critical to the
science, e.g. gravitational wave events, 
gamma-ray bursts, and supernovae.
We have encoded the technique in Python 
on GitHub:
https://github.com/FRBs/astropath.
\end{abstract}


\section{Introduction} \label{sec:intro}

Transient phenomena offer terrific potential to
explore astrophysical processes on the smallest scales
and in the most extreme conditions.  This includes spectacular
explosions, intense magnetic fields, gravity in the 
strong limit, and the structure of dark matter.
Given the very short time-scales, the majority of these 
events are linked to compact objects, e.g.\ neutron stars
and black holes \citep[e.g.][]{Fishman95,GalYam19,Cordes19}, 
and therefore they give unique insight to the
processes that generate and destroy these exotic bodies.

The three-dimensional location of transient sources is a critical aspect influencing the interpretation of their nature, allowing measured properties to be translated into absolute energetics and determining the nature of their environment.  While many transient sources can be reasonably well-localised on the sky, depending on the nature of the discovery instrument (and any sufficiently prompt follow-up), the third dimension of distance is often challenging to obtain.
For some transient phenomena --- supernovae, the afterglows
of gamma-ray bursts (GRBs) ---
spectra of their electromagnetic emission 
can be used to identify their redshift \citep[e.g.][]{blondin2007,Fynbo09}.
Many other transients, however, encode no direct measure of
the source redshift.  This includes the enigmatic fast
radio bursts (FRBs) whose dispersion measures (DMs)
imply a cosmological origin \citep{Lorimer07}, yet
do not provide a precise redshift estimate 
\citep[e.g.][]{McQuinn14,xyz19}. 
Another example includes short-duration GRBs whose
afterglows are too faint to record a high S/N
spectrum for precise redshift estimation \citep[e.g.][]{Fong10}.

When a redshift cannot be measured based on the transient source itself, 
a suitable alternative is to associate the transient event with a galaxy and then measure the galaxy redshift
\citep[e.g.][]{Tendulkar17}.  
This presumes, of course,  that the transient is generated in 
(or at least near) a galaxy -- a reasonable assumption for compact objects
which, aside from exotic and unproven phenomena such as
primordial black holes,
are born in the dense regions of galaxies.
Some progenitors of transient sources may travel considerable distances from their birth sites, for instance via `kicks' during
formation events or disruptions of stellar multiples.
For most such cases, however, offsets of up to tens kpc or several arcseconds on the sky
can be expected for distant events \citep[e.g.][]{Fong13}. 

The process of associating a transient
to its host galaxy is a non-trivial exercise.  Primarily, this is influenced by the uncertainty in the transient localization combined with
the relatively high surface density of galaxies on the
sky, meaning that the allowed region for the transient site can potentially overlap with multiple galaxies.  The initially unknown origins of many such phenomena --- 
including the transients of our interest, FRBs  --- further compounds the problem by introducing an uncertainty in the characteristic offset of a source from the centre of a galaxy.

To date, host associations for transient sources have focused
on the probability of chance association.
\cite{Bloom02} introduced the concept of a chance
probability \pchance\ 
to ascertain the likelihood that a given galaxy was
a coincident association to a transient event.
By inference, a galaxy with a very low \pchance\ value 
might be considered the host,
while galaxies with $\mpchance \sim 1$ may be disfavored
as unrelated sources.
\cite{Tunnicliffe14} advanced this approach by allowing
for galaxy-galaxy clustering which modifies the random 
incidence from a strictly Poisson process.
Most recently, \cite{EftekhariBerger2017} discussed
this approach in the context of FRBs, emphasizing the
need for sub-arcsecond localizations for secure
host galaxy associations.

While we are primarily motivated by FRB science, 
the formalism introduced here is general, and
we identify obvious applications to other
transients, e.g.\ GRBs and GW events.
Our guiding principles for the development of a new
methodology to assess host associations are to:

\begin{itemize}
 \item Be driven by simple observables (which are defined below).
 \item Assign a posterior probability to every candidate galaxy in consideration.
 \item Develop an extendable framework that can evolve as the 
 field matures.  This includes incorporation of
 additional observational constraints and priors.
 \item Accommodate transients both in the local 
 (hundred Mpc) and very distant universe. 
 \item Allow for insufficient data, e.g.\ the non-detection
 of the host galaxy due to imaging depth.
\end{itemize}

In the following, we strive to limit the analysis to
these easily attainable, direct observables:

\begin{enumerate}
    \item The transient localization (RA, DEC, uncertainty ellipse): 
    \rafrb, \decfrb, \eefrb.
    \item The apparent magnitudes of the galaxy candidates:
    $\mgmag_i$.
    \item The candidate galaxy coordinates:
    \ragal, \decgal.
    \item The angular size of the
    galaxy candidates: \galhalflight.
\end{enumerate}
Future work will consider additional observables (e.g.\ the FRB DM) 
and also priors based on ``secure'' host associations.

This paper introducing the probabilistic association of
transients to hosts (PATH) is organized as follows:
section~\ref{sec:chance} briefly reviews the historical approaches
to associations, section~\ref{sec:bayesian} introduces our
new method, section~\ref{sec:priors} defines the priors adopted
for our FRB analysis, section~\ref{sec:sim} presents analysis
of simulated transients, and section~\ref{sec:results} 
applies the formalism to real FRBs.
Throughout we adopt the \cite{Planck15} cosmology as
encoded in {\sc astropy}.

\begin{figure}[!ht]
\centering
    \includegraphics[width=0.5\textwidth]{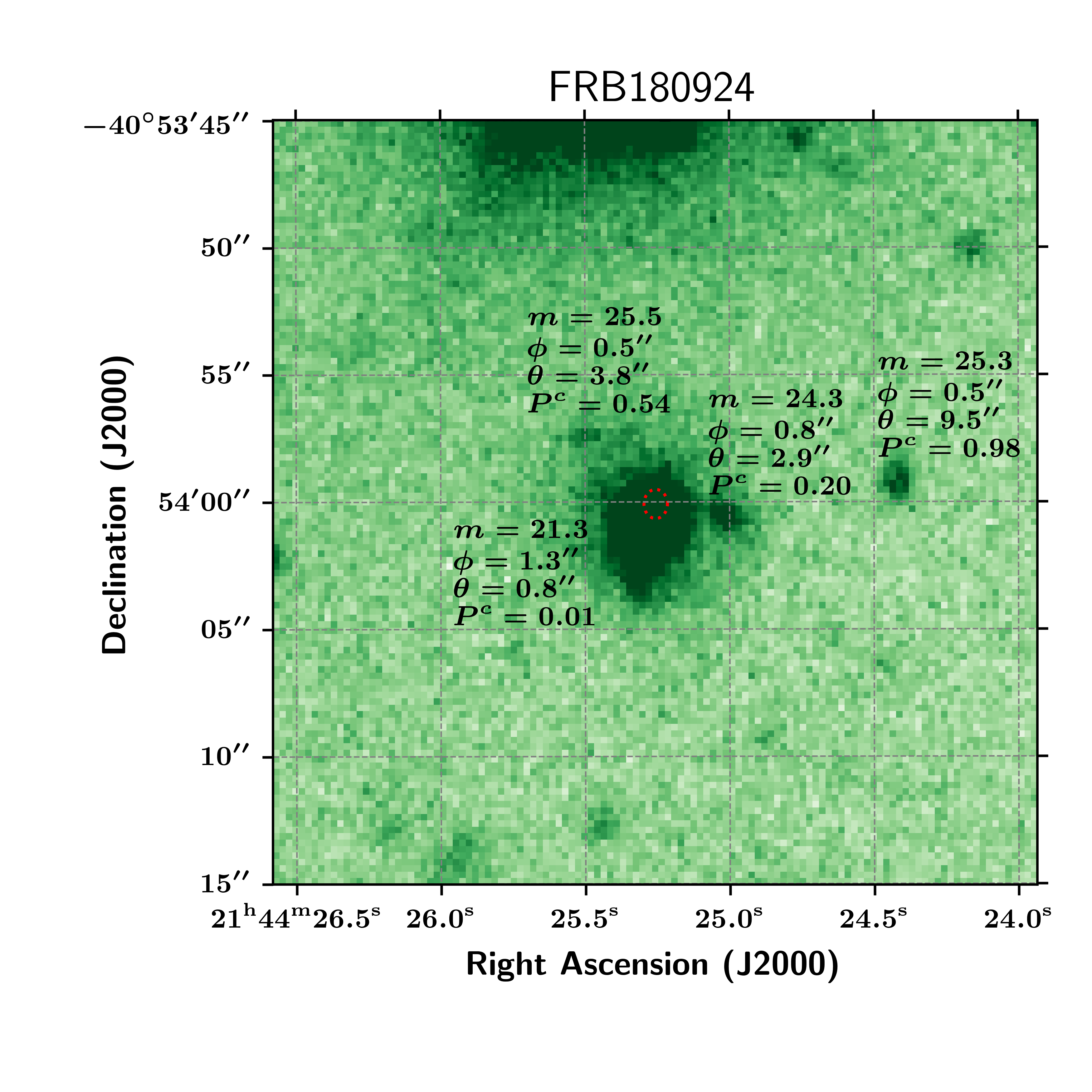}
    \caption{
    Cutout $g$-band image (VLT/FORS2) centered on FRB~180924
    (small red circle indicates the $5\sigma$ localization).  There are four extended sources with
    separation $\theta < 10''$ marked as candidates for the
    host galaxy of the FRB.
    These are labeled (just above them, 
    except for the brightest)
    by their $\theta$, angular size
    \halflight, apparent magnitude $m$, and the chance
    probability \pchance\ of an association.
    }
	\label{fig:chance}
\end{figure}

\section{Historical: Chance Probability} \label{sec:chance}

A standard approach to associating transients to their
host galaxies is through assessing the chance 
probabilities \pchance\ of galaxies 
being located close to the transient position.
We reintroduce the formalism for evaluating \pchance\ here,
propose a new variant, and comment further 
on its application and limitations.

\subsection{Formalism}

Figure~\ref{fig:chance} shows a VLT/FORS2 $g$-band
cutout image ($30'' \times 30''$)
of the field surrounding FRB~180924 with its localization
marked by a red circle.
As described by \cite{Bannister19}, this localization lies
$\approx 1''$ from the centroid of galaxy 
DES~J214425.25$-$405400.81 and 
within $5''$ of two additional galaxies.
At their redshifts \citep{Bannister19}, 
the galaxies all have projected separations of 
less than 40\,kpc, i.e.\ 
separations less than the estimated radii of their
halos.
Therefore, while one may be predisposed to assign 
FRB~180924 to the brighter and closer galaxy, one
should also entertain the possibility that the FRB occurred in the 
stellar halo of one of the others.

The chance probability approach is powerfully simple:
estimate the Poisson probability of finding one or more 
galaxies as bright or brighter within an effective
search area $A$ around the FRB, with $A$ determined from the 
angular size \halflight, 
the separation \sep, and the 
FRB localization uncertainty \sigmafrb.
Namely, one defines the probability of a chance
coincidence 

\begin{equation}
  \mpchance(\mtheff, \mgmag)  = 1 - \exp(-\bar N) \;\;\; ,
\label{eqn:chance}
\end{equation}
where $\bar N$ is the average number of sources
in $A$. It is given by

\begin{equation}
\bar N(\mtheff, \mgmag_i)  = \pi \mtheff^2 \, \Sigma(\mgmag_i) \;\;\; ,
\label{eqn:avgN}
\end{equation}
where \numden\ is the angular surface density of 
galaxies on the sky with magnitude $\mgmag \le \mgmag_i$,
and \theff\ is the effective search radius ($A = \pi \mtheff^2$).
For the former, we adopt the galaxy number count distribution
of \cite{driver16},
while the latter quantity bears some arbitrariness.

Previous works have considered several definitions
for \theff.  We introduce yet another more conservative
one here: the quadrature sum of all three
angular quantities with semi-arbitrary weightings,

\begin{equation}
\mtheff = \sqrt{4 \msigmafrb^2 + \theta^2 + 4 \mhalflight^2} 
\;\; .
\label{eqn:theta_eff}
\end{equation}

Adopting these, we estimate $\mpchance = 0.01$ for 
DES~J214425.25$-$405400.81 and $\mpchance > 0.1$ for
the other sources owing to their fainter magnitudes and
larger angular separations.  
From this perspective, the probability of a chance
association is nearly negligible for DES~J214425.25$-$405400.81 
and sufficiently large for the other sources that one is inclined
to favor the former.
However, this technique cannot assign a likelihood to
this association, nor even a relative assessment of one
source over another.
Yet worse, if two or more candidates have low \pchance\ 
\citep[e.g. FRB~181112;][]{Prochaska19b}, one has no means
to favor one over the other.
Last, the \pchance\ formalism does not naturally allow one
to introduce additional observational measures
as evidence (e.g.\ DM).
Together, these considerations motivate our development
of a full probabilistic treatment.

\subsection{Nuisances and Nuances}

There are several aspects of the \pchance\ analysis that require
further definition.
For completeness, we describe these here 
although \pchance\ does not formally enter into the
new formalism.
First, one requires measurements of the galaxy centroids
and angular size.
We advocate a non-parametric approach owing to the complexity
of galaxy morphology.
In the following, we adopt the centroiding algorithm 
encoded in the \photutils\ software package,
and use the \texttt {semimajor\_axis\_sigma} 
parameter to estimate the angular size. 
In the following, we will refer it as \aimage\ 
as it matches the definition of that parameter
in the more widely used {\sc sextractor} package.

Another issue is Galactic extinction.
FRBs are detected across the sky including on sightlines that
show large Galactic extinction ($\mebv > 0.1$\,mag).
In contrast, the number count analyses have intentionally
been derived from high-latitude fields with low
Galactic extinction.  Therefore, the apparent magnitudes
of the galaxy candidates should be corrected for 
Galactic extinction prior to the probability estimation.

We provide the following set of recommendations to optimize the detection and characterization of both faint and bright sources. Namely, we recommend observations in the $r$-band, which provide a trade-off between the mapping of stellar content and extinction. We further recommend an image depth of $m_r = 25.5$ ($5\sigma$), corresponding to the limiting magnitude for spectroscopy, and sufficient for probing $0.01 \ L^*$ galaxies at $z\sim 0.5$. Finally, given the subarcsecond accuracy of many FRB localizations, we recommend better than $1''$ seeing, and a $>1'$ field of view for background estimation. 

Similarly, source detection should employ a non-parametric approach to accommodate galaxies of varying morphologies. For consistency with this work, we recommend use of the {\texttt{a\_image}} parameter to estimate the angular size of sources, and note that all galaxies should be corrected for Galactic extinction prior to applying the Bayesian formalism.

The formalism of eq.(\ref{eqn:chance}) and (\ref{eqn:avgN}) ignores galaxy clustering. Since matter in the Universe is not uniformly distributed, the probability of observing either no galaxies, or a large number of them, in proximity to a random direction is enhanced compared to the probability of observing one or a few. \citet{Tunnicliffe14} show that including clustering decreases the probability of a nearby random galaxy by 25--50\% in the case of a random direction. Our primary concern however is not whether or not all the observed galaxies are merely chance coincidences, but rather which of the observed galaxies is the true host.
Clustering is discussed further in this context in section~\ref{sec:candidate_priors}. For now, we remark that given that FRBs truly are associated with galaxies, clustering acts to increase, not decrease, the probability of a chance association, since the FRB observation has preferentially selected a direction of the Universe in which there is a cluster of matter.

Furthermore, in this work we ignore for the sake of simplicity the ellipticity of the prospective host galaxies. 
Our method will be readily adaptable however to such ellipticity, or indeed arbitrarily complex functions, since the approach described below does not rely on any particular functional forms.

\section{Probabilistic Approach} \label{sec:bayesian}

Association of transients
to galaxies is not like the usual cross-identification for which probabilistic methods have been in place for over a decade \citep{budavari_szalay_2008}. 
Matching stars and galaxies typically involves asking whether a set of detections (across separate exposures, instruments, telescopes) are of the same celestial object. 
If they are, their true (latent) directions would have to be the same; see more in the review by \citet{budavari_loredo_2015}. 

In strong contrast with that, FRBs are presumed
to simply originate from within (or at least near to)
galaxies, hence their true direction should not be required to coincide with the center of a galaxy. While this is admittedly a small difference for the faintest galaxies, resolved extragalactic source are expected to yield different results.
Here we consider a general scenario where the shape of galaxies can be incorporated along with a geometric model about from where FRBs would originate within 
or around galaxies.

\subsection{General Formalism} 

Since FRBs are sparse on the sky, we can study them separately, which also simplifies the following description of our approach. 
Let us consider a catalog of galaxies across the entire sky and a single FRB that either belongs to one of the many catalog objects (its host galaxy) or it does not, i.e.\ its host is not detected or not included in the catalog. 
If $U$ is the event that the FRB's host is unseen, and $O_i$ is the event that the FRB is from galaxy~$i$, their prior probabilities must add up to one,
\begin{equation}
\mPU +  \sum\limits_i \mPOi = 1 \;\;\; ,
\label{eqn:norm}
\end{equation}
as there are no other possibilities.
The single scalar quantity $P(U)$ encodes all the complications that arise from the difference in the radial selection functions of the catalog and the FRB instruments. For now, we assume its value to be known, but note that it could be inferred in a hierarchical fashion when considering multiple FRBs.
Also, one could assume a uniform prior for all observable $O_i$ as the simplest possible scenario that essentially ignores any additional information about the galaxies, e.g., magnitude, color, redshift. We leave such considerations to a future work.

Given a vector $x$ representing all measured properties of the detected FRB, we ask what the posterior probabilities $\mPUx$ and $\mPOix$ are for all $i$. 
Using Bayes' rule, the unseen posterior is
\begin{equation}
\mPUx = \frac{\mPU\,p(x|U)}{p(x)} \;\;\; ,
\label{eqn:U}
\end{equation}
where $p(x|U)$ is the probability density of the FRB properties given the host is unseen. From hereon, we consider $x$ to represent only the measured FRB direction.
Without constraints, $x$ could be anywhere on the sky, hence it is natural to assume a uniform (isotropic) distribution with a value of $1\big/4\pi$.
Similarly, the posterior probability for object $i$ is
\begin{equation}
\mPOix = \frac{\mPOi \, \mpxOi}{p(x)}  \;\;\; ,
\label{eqn:bayes}
\end{equation}
where $\mpxOi$ is the probability density function (PDF) of $x$ given that the FRB comes from galaxy $i$. 
With data $x$, this is the marginal likelihood of $O_i$, which includes the galaxy geometry and the uncertainty of the FRB direction. 
This key component of the approach is discussed in depth in the next paragraph.
The normalizing constant must be
\begin{equation}
    p(x) = \mPU\,p(x|U) + \smm_i \mPOi\,\mpxOi \;\; 
\label{eqn:denom}
\end{equation}
to guarantee that these posteriors also add up to one,
\begin{equation}
     \mPUx + \smm_i \mPOix  = 1 \;\; .
\label{eqn:posterior_norm}
\end{equation}

\subsection{Marginal Likelihoods}
\label{sec:marg_like}

Let the 3-D unit vector $\omega$ represent the 
{\it true} and unknown direction of the FRB on the sky.
Given that it comes from a particular galaxy, the direction $\omega$ has to point somewhere near the host.
The function \poffset\ captures the physical and geometric model for the FRBs specific to galaxy $i$, e.g., taking into account its type, distance, orientation, etc. 

The observed FRB direction $x$ is a measurement of $\omega$ with known uncertainty, represented by the localization error function, $L(x-\omega)$. 
Given \poffset,  $\mpxOi$ is calculated by integrating over the $\omega$ model directions to obtain the marginalized likelihood of the association hypothesis $O_i$, 
\begin{equation}
\mpxOi = \int\!\!d\omega\, \mpoffset \, L(x-\omega) \ ,
\label{eqn:pxO}
\end{equation}
which now accounts for all possibilities in various FRB origins as well as the astrometric uncertainty in the measurement.
If a galaxy is unresolved, $p(\omega|O_i)$ may become the Dirac-$\delta$, and $\mpxOi$ is just the astrometric uncertainty, as it would be the case for matching point sources.
Calculating the above quantity for all $i$ 
completes the framework, which now provides posterior probabilities via
eq.~\ref{eqn:bayes}.

\subsection{Limited Field of View}
\label{sec:limited_field_of_view}

Previously we assumed a galaxy catalog over the entire sky, but catalogs typically have more limited footprints.
It is interesting to think about a scenario when the field of view $\Omega$ is smaller but still large enough not to miss any possible counterparts to the FRB in question.
Intuitively, galaxies very far away from the FRB should not have any effect on the association analysis.

Smaller sky coverage would mean fewer observed galaxies \Nc, which in turn affect the \POi\ priors as there are fewer galaxies to choose from. Going with the previous uniform assumption, eq.(\ref{eqn:norm}) implies
\begin{equation}
    \mPOi = \frac{1-P(U)}{\mNc} \;\; ,
\label{eqn:equalPO}
\end{equation}
which captures the dependence.

Looking now back at Bayes' rule in eq.(\ref{eqn:U}), 
the sky coverage seems to affect the $\mPOix$ posteriors, too.
Fortunately, this is not the case. 
The denominator $p(x)$ changes in accord due to the scaling in $p(x|U)$, which is uniform over the field of view,
\begin{equation}
    p(x|U) = \frac{\boldmath{1}_{\Omega}(x)}{\Omega}
\end{equation}
where the $\boldmath{1}_{\Omega}(x)$ is the indicator function that takes the value 1 if $x$ is within the field of view and 0 otherwise, and the denominator $\Omega$ normalizes the PDF.
The framework not only matches common-sense expectations, but provides for more efficient computation where only candidates within close proximity of FRBs are considered.

Given the plethora of deep, wide-field imaging
surveys, it is possible to consider very large $\Omega$
for each FRB in the analysis.  In practice, however, 
sensible and conservative assumptions for \poffset\ 
will greatly limit the list of viable candidates $O_i$.
To ease the analysis, we adopt for $\Omega$ the union
of the area encompassing all galaxies within 10 half-light
radii and the \frblocal\ FRB localization.
As emphasized in the previous section, it is only necessary to consider a large enough area to be certain to include all possible hosts.

\section{Priors and Assumptions}
\label{sec:priors}

\subsection{Undetected Prior \PU}
\label{sec:PU}

It is difficult to {\it a priori}
assign a prior \PU\ to the probability
that the FRB host is undetected.  Undoubtedly, 
\PU\ is related to the depth of the imaging and the
(unknown) source distance, i.e. redshift\footnote{
Future work may use the observed DM to inform 
the host distance and \PU.}.
In the following, we consider 
an arbitrarily assigned value,
and we advocate a low value,
based on the paucity of our data and the set of
confidently assigned associations reported
to date \citep[e.g.][]{Heintz2020}.
In the analyses that follow, we typically assume
$\mPU = \uguess$ (Occam's razor!) 
and discuss the impacts of increasing it.


\subsection{Candidate priors \PO}
\label{sec:candidate_priors}

For the set of host candidates $O_i$,
absent any assumptions on the distance or the typical
separations of FRBs from their host galaxies,
any galaxy on the sky could be a viable candidate.
Given the plethora of potential models for FRB progenitors
and the limited existing constraints, we are motivated
to consider (for now) simple approaches to the prior for the
galaxy candidates.
The most agnostic approach is to assign an ``identical''
prior to every galaxy in consideration, 
i.e.\ eq.~\ref{eqn:equalPO}.

Inspired by the chance probability calculations
approach described in $\S$~\ref{sec:chance}, 
we introduce an
additional prior based on \pchance. 
Specifically,  we consider a prior that inversely weights by \numden.

\begin{equation}
\mPOi \propto \frac{1}{\minumden} \;\; .
\label{eqn:prior}
\end{equation}
For this ``inverse'' prior, 
brighter candidates have higher prior probability
according to their number density on the sky.
The normalization of these priors is set by 
eq.~\ref{eqn:norm}.
In section~\ref{sec:sim} we also briefly consider
two other priors with 
$\mPOi \sim 1/\minumden \phi$ (inverse1)
and
$\mPOi \sim 1/\minumden \phi^2$ (inverse2).
We adopt the prior in Equation~\ref{eqn:prior}
in part because of its simpler form and also 
because of the results of simulated experiments
(Section~\ref{sec:sim}).

\begin{figure*}[!ht]
\centering
    \includegraphics[width=5.9cm]{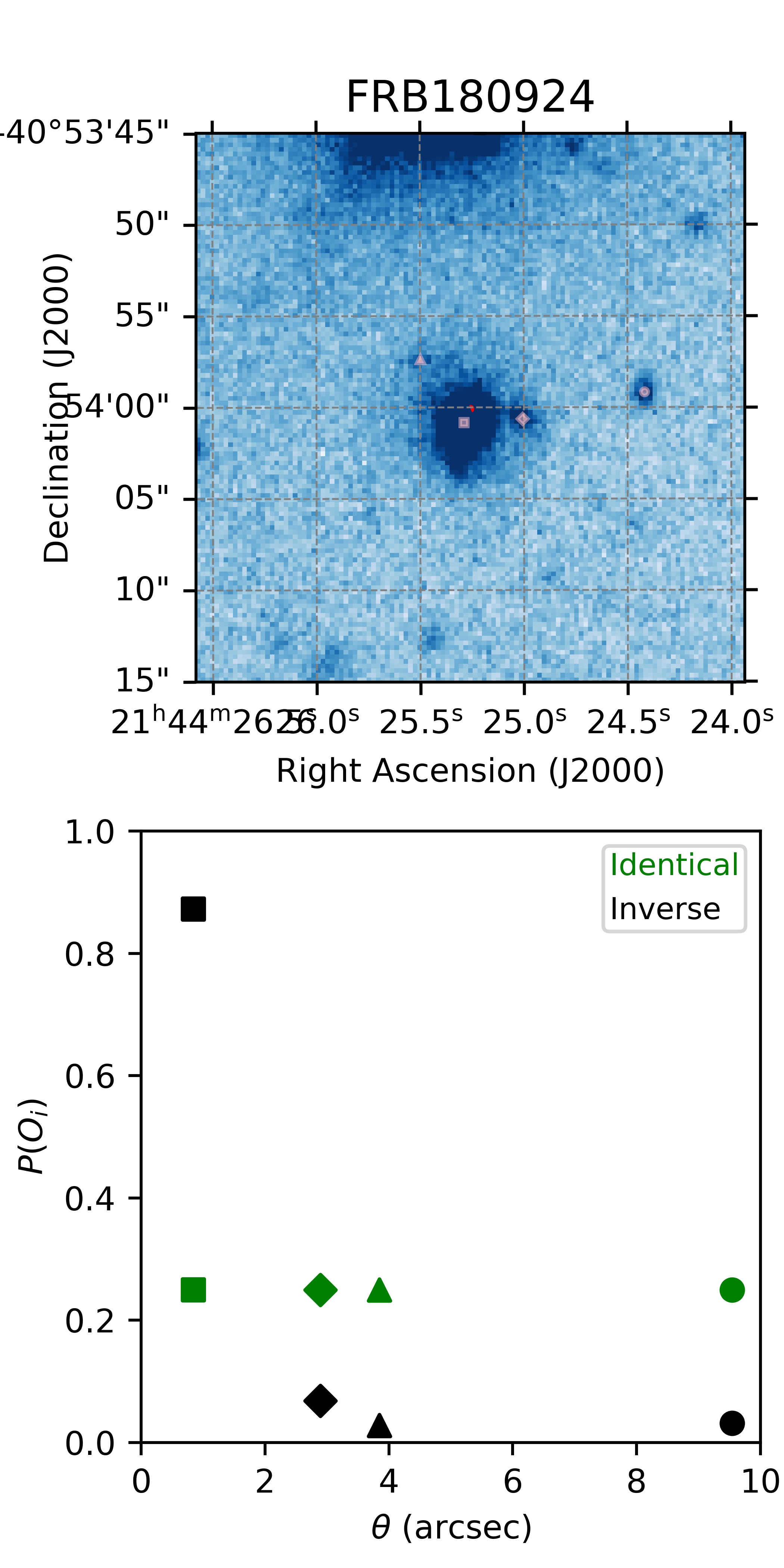}
    \includegraphics[width=5.9cm]{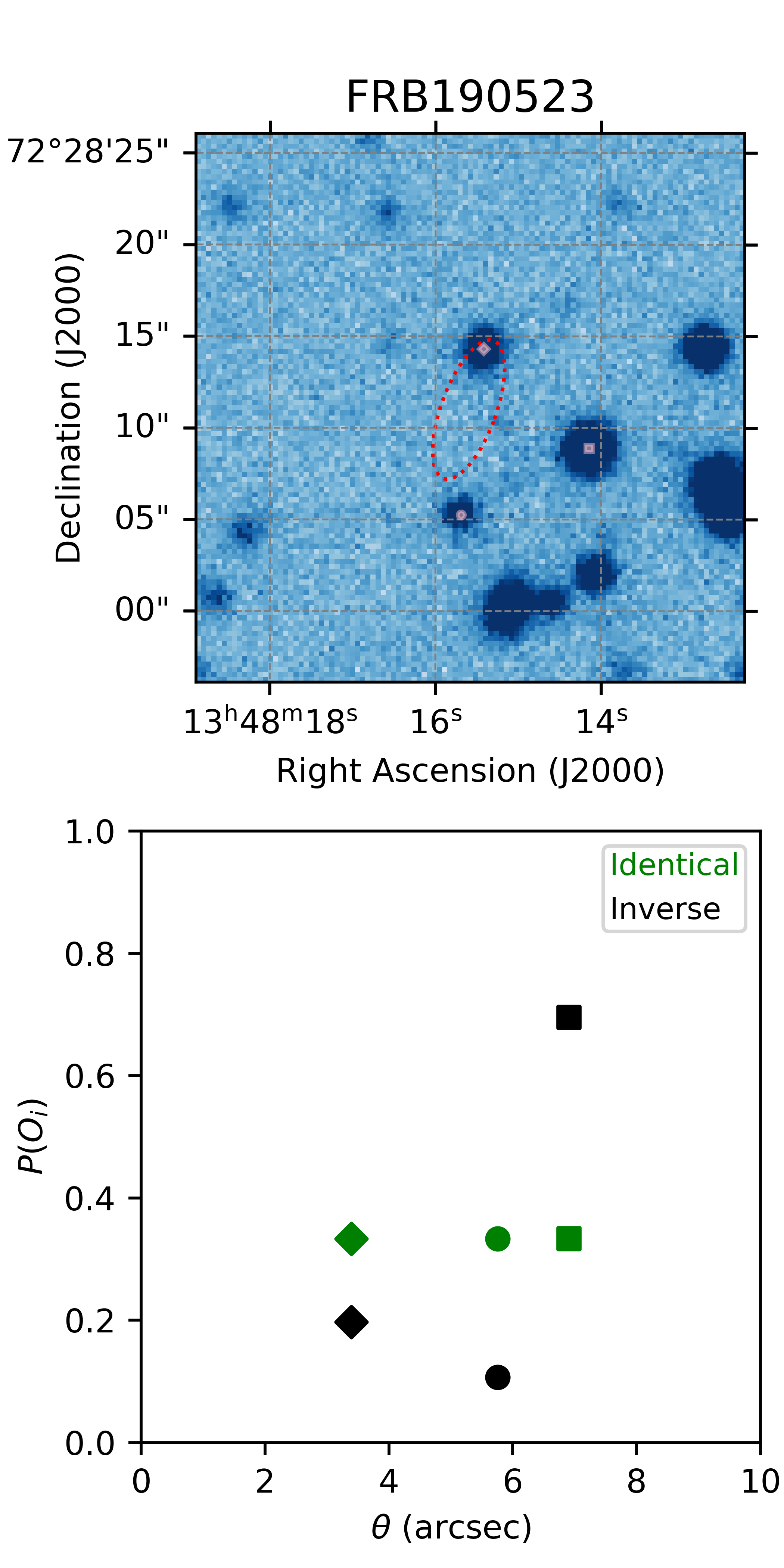}
    \caption{
    Illustration of the two approaches to candidate priors
    \PO\ assumed in this manuscript:
    identical and inverse.  The former assumes an identical
    prior for every galaxy with $\theta_i < 10 \mhalflight$
    or within the \frblocal\ localization error of the FRB.
    The latter adopts the inverse of the
    estimated chance probability \pchance\ 
    of these galaxies which is equivalent to asserting
    the prior is proportional to the integrated probability
    that the other sources are all chance coincidences.
    }
	\label{fig:galaxy_priors}
\end{figure*}

Figure~\ref{fig:galaxy_priors} illustrates 
\POi\ for the two models for two example cases --
(a) FRB~180924 and (b) FRB~190523 --  where
we assume $\mPU = \uguess$.
For this illustration, we have restricted the analysis to galaxies
within $10''$ of the FRB, which captures all of the viable
candidates.  As expected, the inverse priors for
FRB~180924 significantly favor the brighter galaxy closest
to the FRB.  Perhaps less intuitively, the 
inverse priors favor the most distant 
(yet brightest) source near
FRB~190523. This motivates the inclusion of the next ingredient --- the offset function \poffset.

\begin{figure}[!ht]
\centering
    \includegraphics[width=0.47\textwidth]{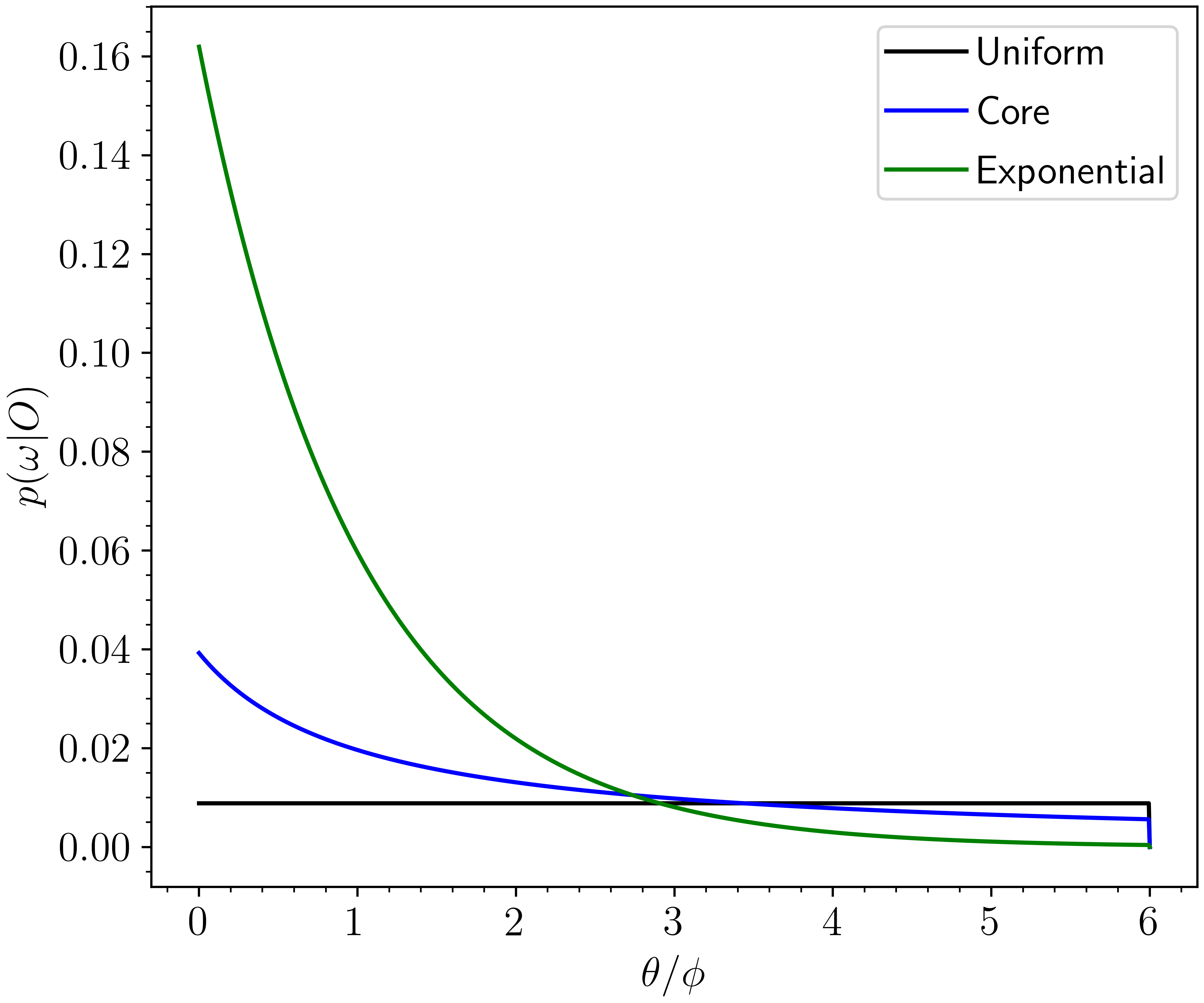}
    \caption{
    Three offset functions \poffset\ considered for the underlying angular distribution of FRBs relative 
    to their host galaxies, normalized by the galaxy's half-light radius \halflight.
    Each is normalized to have identical integrated area to a maximum assumed offset of $\mthmax = 6 \, \mhalflight$. 
    }
	\label{fig:theta_prior}
\end{figure}

\subsection{Offset function \poffset}
\label{sec:poffset}

The \poffset\ function
for the probability of the true angular offset from the 
galaxy is unknown yet required for the analysis. 
In our formalism, we develop priors for \poffset\ based
solely on the angular offset $\theta$ between the galaxy
centroid and $\omega$, and also
normalized by the observed galaxy's angular size
\halflight.  This simultaneously accounts
for galaxies of different intrinsic size and differing
observed size owing to their distance.

As the predominance of models associate FRBs
to stellar sources or compact objects
\citep[AGN at the very centers of galaxies are currently disfavored;][]{Bhandari20a}, one might expect
the FRB events to track the stellar light. 
While we wish to remain largely agnostic to the
underlying distribution of offsets, we are
physically motivated to presume \poffset\ 
decreases with increasing $\theta$.
We assert this despite the fact that geometrical
considerations do favor large $\omega$, e.g.\ 
a model where FRBs occur with identical probability
anywhere in a circular galaxy will have
$\mpoffset \propto \omega$ until one reaches the
``edge'' of the galaxy.
Therefore, a {\it uniform} prior 
$\mpoffset =  1 / \pi \theta_{\rm max}^{2}$
is formally one that assumes FRBs occur 
proportional to the galaxy radius.
The other two models considered are a {\it core} model: 

\begin{equation}
    \mpoffset = 
    \frac{1}{2 \pi \phi^2 [\mthmax/\phi - \log(\mthmax/\phi + 1)]}
    \frac{1}{(\theta/\mhalflight) + 1} \;\;\; ,
\label{eqn:poff_core}
\end{equation}
which implies an approximately $1/r^2$ weighting,
and an {\it exponential} model

\begin{equation}
    \mpoffset = 
    \frac{1}{2 \pi \phi^2 [1 - (1 + \mthmax/\phi) \exp(-\mthmax/\phi)]}
    \exp[-\theta/\mhalflight] \;\;\; ,
\label{eqn:poff_exp}
\end{equation}
which assumes an underlying exponential distribution.
All of these functions are normalized to 
unity when integrating to $\theta_{\rm max}$, ignoring
the curvature of the sky.

For all of the \poffset\ priors, we assert a maximum
offset $\mthmax = 6 \mhalflight$; this is especially
important for the uniform prior.  
This value is arbitrary and was chosen to 
be large enough to accommodate prevailing models of
FRBs without being too conservative.

Applying an arbitrary 
cutoff to the exponential distribution is not strictly
necessary, but we keep it for simplicity and consistency, 
and demonstrate in Section~\ref{sec:add_priors} that the 
results for an exponential distribution are insensitive 
to this choice.

Figure~\ref{fig:theta_prior} shows the offset functions,
normalized to have the same total probability.
Clearly the {\it exponential} model favors FRBs located in the inner
regions of galaxies.
For comparison to the offset distributions of known transients, we note that long GRBs appear highly concentrated in the inner regions of their hosts relative to Type Ib/c and IIn supernovae, which occur preferentially near their host half-light radii \citep{Lunnan15,Blanchard16}. Conversely, short GRBs exhibit significant offsets from their host centers, indicative of progenitors born in compact object mergers \citep{Fong13}.

In practice,  we treat the galaxies as ``round'',
i.e.\ ignoring for now any ellipticity.
Future works will advance this aspect.

\section{Simulations}
\label{sec:sim}
To explore the formalism introduced here, we have generated
Monte Carlo simulations designed to faithfully reproduce
the FRB experiment.  We describe this first and then detail
the results.

\begin{figure}[!ht]
\centering
    \includegraphics[width=8.5cm]{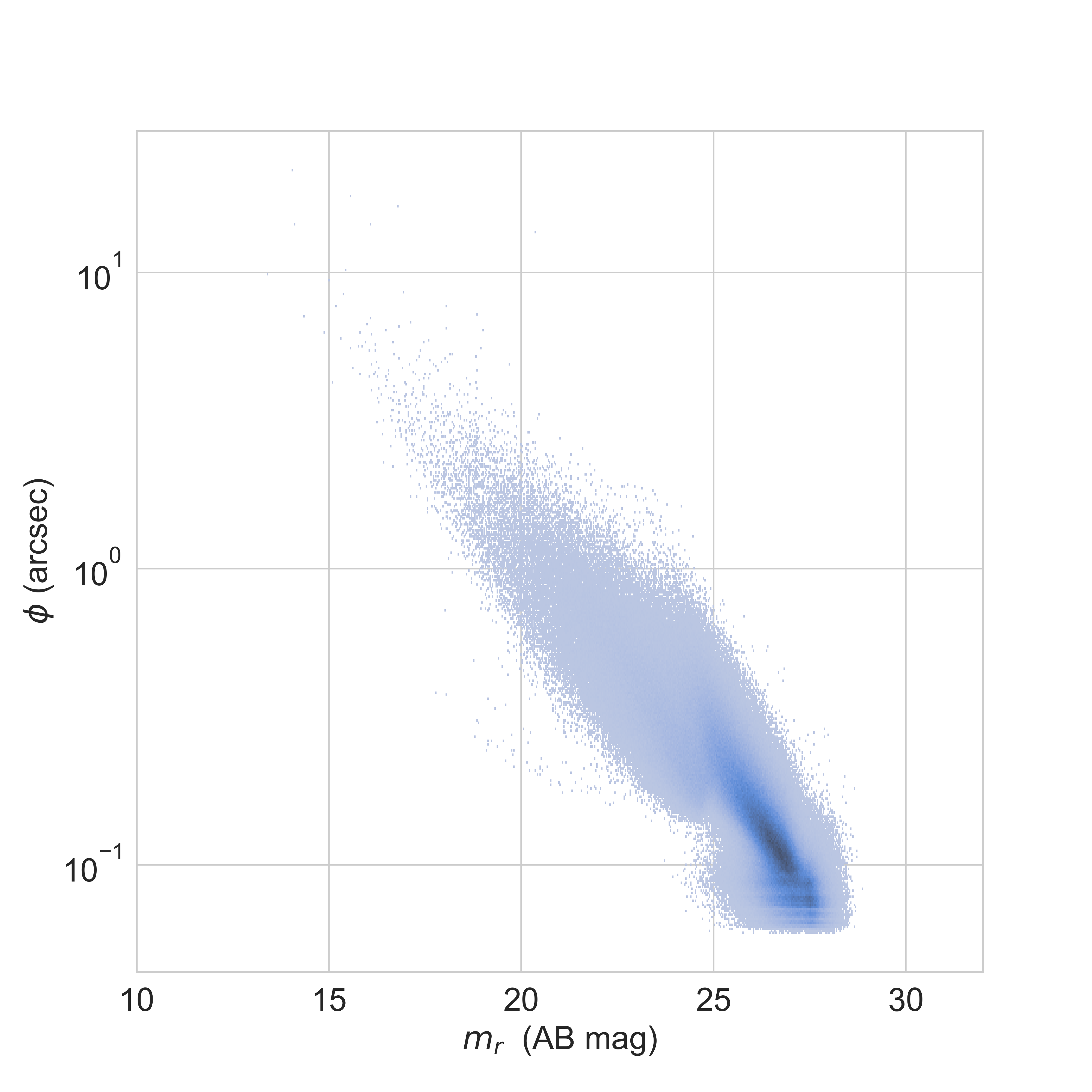}
    \caption{
    Distribution of angular sizes (\aimage\ parameter)
    versus apparent magnitude $m_r$ for all of the galaxies
    in the COSMOS catalog \citep{Scoville2007cosmoshst}.
    }
	\label{fig:cosmos}
\end{figure}

\subsection{Sandboxes}
\label{sec:sb}

Our Monte Carlo approach leverages the public catalog of 
the {\it HST}/COSMOS field \citep{Scoville2007cosmos, Scoville2007cosmoshst}
which provides over 1~million sources at high spatial
resolution and faint fluxes (median 
AB magnitude $m_r \approx 25.9$).
These galaxies are distributed over an $\approx 1$\,sq.\,deg
of sky which captures a great variety of distributions
but not the most extreme events 
(e.g.\ $z<0.5$ galaxy clusters, nearby galaxies). 
Figure~\ref{fig:cosmos} shows the angular size
(\aimage) and apparent magnitudes for the catalog,
restricted to the sources labeled as galaxies.

We generate a series Monte Carlo realizations of FRBs
(referred to as a sandboxes, or SBs), using the following
recipe:

\begin{enumerate}
    \item Define the true distribution of FRBs from their
    host galaxies \poffset and magnitude $m_r$.
    \item Define a sample of potential host galaxies based
    on $m_r$.
    \item Define the distribution of localization errors for FRBs (\sigmafrb).
    \item Draw $N_{\rm FRB}$ galaxies from the parent
    sample of potential host galaxies without duplicates.
    \item Set the true FRB positions according to \poffset.
    \item Offset the FRBs to an observed coordinate according
    to \sigmafrb.
    \item Consider catalogue galaxies within $30''$ to represent an image.
\end{enumerate}

For sandbox 5 (SB-5), 10\% of the FRBs were randomly placed in the COSMOS field, i.e without a host galaxy. We plan to use this sandbox 
to evaluate the performance of the framework when the host galaxies are unseen. 
Also, as COSMOS is a deep survey, we generate a magnitude-limited 
catalog of galaxies for each sandbox (last column of Table~\ref{tab:sb}) on which we run the Bayesian framework. In the following sub-section, we 
discuss results for five sandboxes
(focusing primarily on SB-1)
with the parameters described
in Table~\ref{tab:sb}.

\begin{deluxetable}{cccccccc}
\tablewidth{0pc}
\tablecaption{SandBoxes\label{tab:sb}}
\tabletypesize{\footnotesize}
\tablehead{\colhead{Label}  
& \colhead{\poffset} & \colhead{$N_{\rm FRB}$} 
& \colhead{Sample}
& \colhead{\sigmafrb}
& \colhead{Catalog Filter}
\\
& & & & ($''$) 
} 
\startdata 
SB-1 & $\mathcal{U}(0, 2\mhalflight)$  & 100,000 
   & - 
   & 1 & - \\
SB-2 & $\mathcal{U}(0, 2\mhalflight)$ & 46,699 & $m_r=[20, 23]$ 
  & $\mathcal{U}(0.1, 1)$ & $m_r \le 23$\\
SB-3 & {\it core} & 46,699 & $m_r=[20, 23]$ 
  & $\mathcal{U}(0.1, 1)$ & $m_r \le 23$\\
SB-4 & {\it exponential}  & 46,699 
  & $m_r=[20, 23]$ & $\mathcal{U}(0.1, 1)$ & $m_r \le 23$\\
SB-5\tablenotemark{a} & $\mathcal{U}(0, 2\mhalflight)$  & 50,000 
  & $m_r=[20, 25]$ & $\mathcal{U}(0.1, 1)$ & $m_r \le 25$ \\
\hline 
\enddata 
\tablenotetext{a}{See text for details regarding the FRB selection for this sandbox.}
\end{deluxetable}

\startlongtable
\begin{deluxetable}{cccccccccccccccc}
\tablewidth{0pc}
\tablecaption{Sandbox Analysis\label{tab:sandbox}}
\tabletypesize{\footnotesize}
\tablehead{\colhead{Sandbox}  
& \colhead{\PO} & \colhead{\PU} & \colhead{\poffset}
& \colhead{\thmax/\halflight}
& \colhead{f(T+secure)} & \colhead{TP} 
\\} 
\startdata 
SB-1&inverse& 0&exp& 6& 0.33& 0.96\\ 
SB-1&inverse1& 0&exp& 6& 0.30& 0.99\\ 
SB-1&inverse2& 0&exp& 6& 0.32& 1.00\\ 
SB-1&identical& 0&uniform& 6& 0.22& 1.00\\ 
SB-1&inverse& 0.05&exp& 6& 0.24& 0.96\\ 
SB-2&inverse& 0&uniform& 2& 0.86& 1.00\\ 
SB-3&inverse& 0&core& 6& 0.58& 1.00\\ 
SB-4&inverse& 0&exp& 6& 0.68& 0.99\\ 
SB-5&inverse& 0.10&exp& 6& 0.58& 0.98\\ 
\hline 
\enddata 
\end{deluxetable}

\subsection{Analysis and Results}
\label{sec:sb_ar}

We now analyze the sandboxes listed in Table~\ref{tab:sb}
with a variety of priors and assumed \poffset\ functions
(that generally do not match the true \poffset).
Table~\ref{tab:sandbox} lists the various priors assumed
for each analysis performed.

\begin{figure}[!ht]
\centering
    \includegraphics[width=8.5cm]{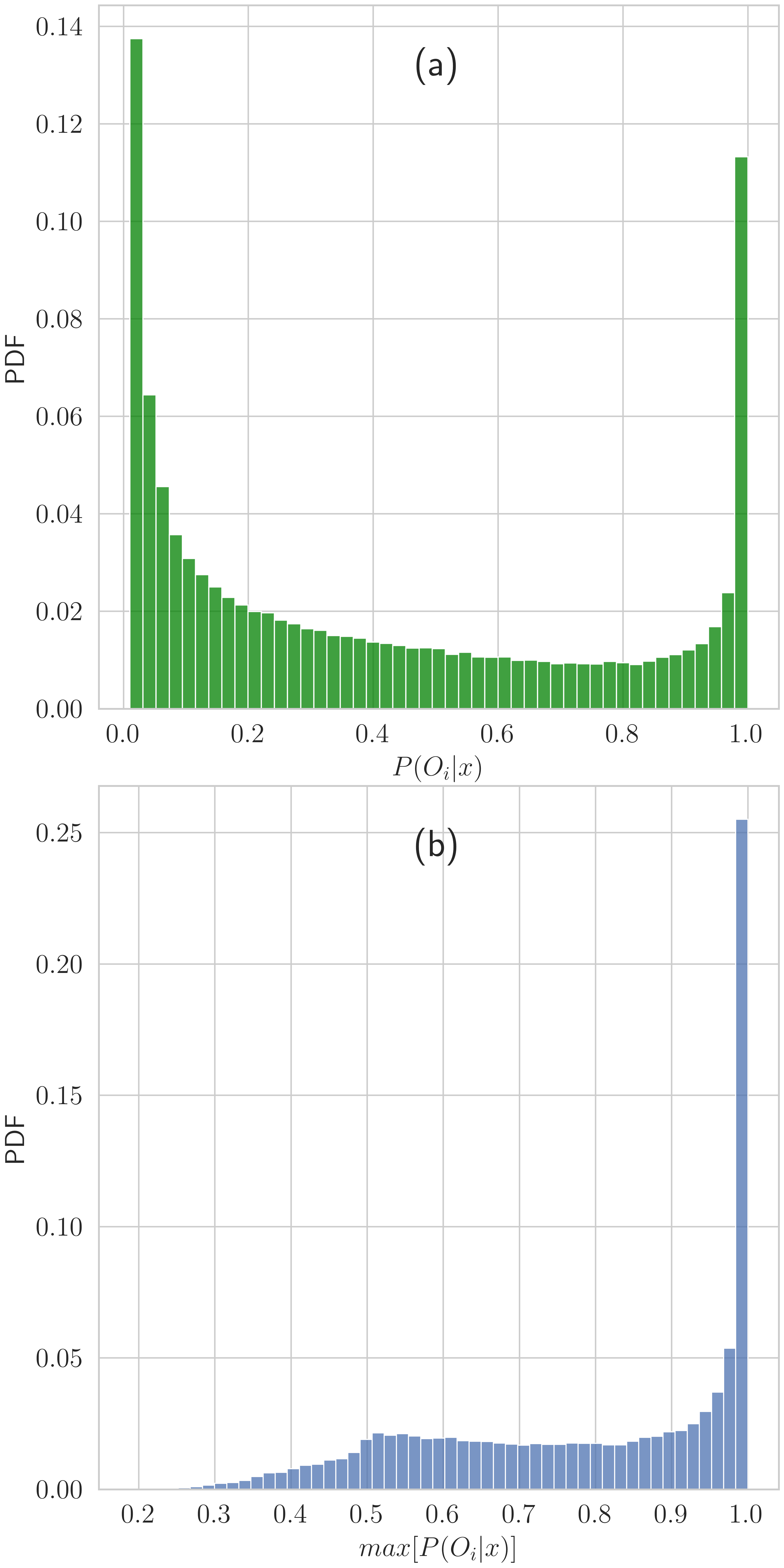}
    \caption{
    Analysis of SB-1:
    (a) PDF for the posterior probabilities for all of the candidates
    with $\mPOix > 0.01$.  The PDF is nicely bi-modal with 
    96\%\ having $\mPOix < 10\%$ or  $\mPOix > 90\%$. 
    (b) PDF of the maximum posterior for the 100,000 simulated FRBs.
    We find $\approx \simsecure \%$ of the sources has 
    $\mPOix > \POvsec$, which we define as secure.
    }
	\label{fig:sb_post}
\end{figure}

Figure~\ref{fig:sb_post}a shows the posteriors for the
candidates using the
fiducial sandbox (SB-1) and
listed in row~1 of Table~\ref{tab:sandbox} 
(also referred to as the \adopted\ prior set;
Table~\ref{tab:priors}).
Since most of the
candidates defined in step 7 are very far
from the offset FRB position, we restrict results
to the $\approx 8\%$ of 
candidates with $\mPOix > 0.01$. 
This distribution is multi-modal, with the overwhelming
majority of recovered $\mPOix \approx 0$ corresponding
to unassociated galaxies and another
peak at $\mPOix \approx 1$ corresponding to secure
associations.
Figure~\ref{fig:sb_post}b shows the posterior
value for the most probable candidate for each of
the 100,000 FRBs.
For this model and analysis, $\approx \simsecure \%$ of
the FRBs have a high probability,
$\mPOix > \POvsec$ which we adopt as a ``secure''
association. Adopting such an arbitrary value
to define `secure' is useful for including/excluding
candidates for subsequent analyses that rely on
knowing the correct FRB host, e.g.\ that by
\citet{Macquart20}. However, we emphasise that
it is in-general better to consider all
host associations as uncertain, with different
levels of certainty according to the obtained
posteriors.

\begin{figure}[!ht]
\centering
    \includegraphics[width=8.5cm]{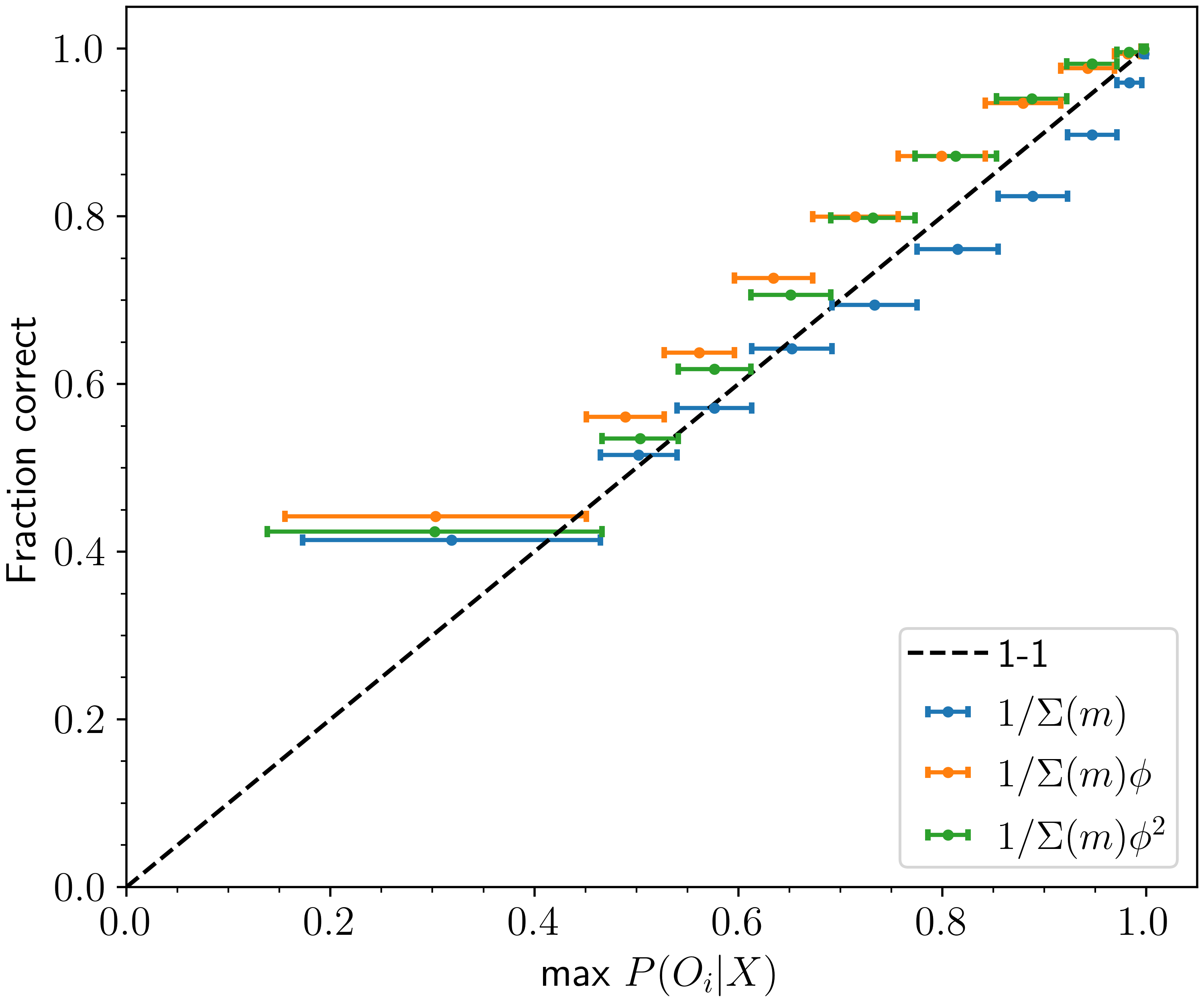}
    \caption{
    The points are equal number (10,000) bins of FRBs
    according to the maximum posterior probability 
    of their host candidates for SB-1.  
    For each set of 10,000
    we determined the fraction of correct assignments
    if one adopts the candidate with max~\POix\ as the
    host.  The one-to-one line indicates a perfectly
    calibrated algorithm.  
    The colors indicate different choices for the
    prior \PO: 
    (blue) $\mPO \propto 1/\Sigma(m)$, inverse;
    (orange) $\mPO \propto 1/\Sigma(m)\mhalflight$, inverse1;
    (green) $\mPO \propto 1/\Sigma(m)\mhalflight^2$, inverse2.
    Our adopted prior (inverse) is relatively well normalized
    in that \POix\ yields an accurate estimate of the
    fraction of FRBs correctly assigned to their host galaxy.
    }
	\label{fig:sb_priors}
\end{figure}

Figure~\ref{fig:sb_priors} evaluates, in
10~bins of equal number of FRBs,
the maximum \POix\ assigned to a candidate
for each FRB and the percentage of correct
associations assuming this is the host.
The different colors indicate 
different choices for \PO.
Our adopted ``inverse'' prior appears well calibrated
in that \POix\ yields an accurate estimate of the
fraction of FRBs correctly assigned to their host galaxy.

\begin{figure}[!ht]
\centering
    \includegraphics[width=8.0cm]{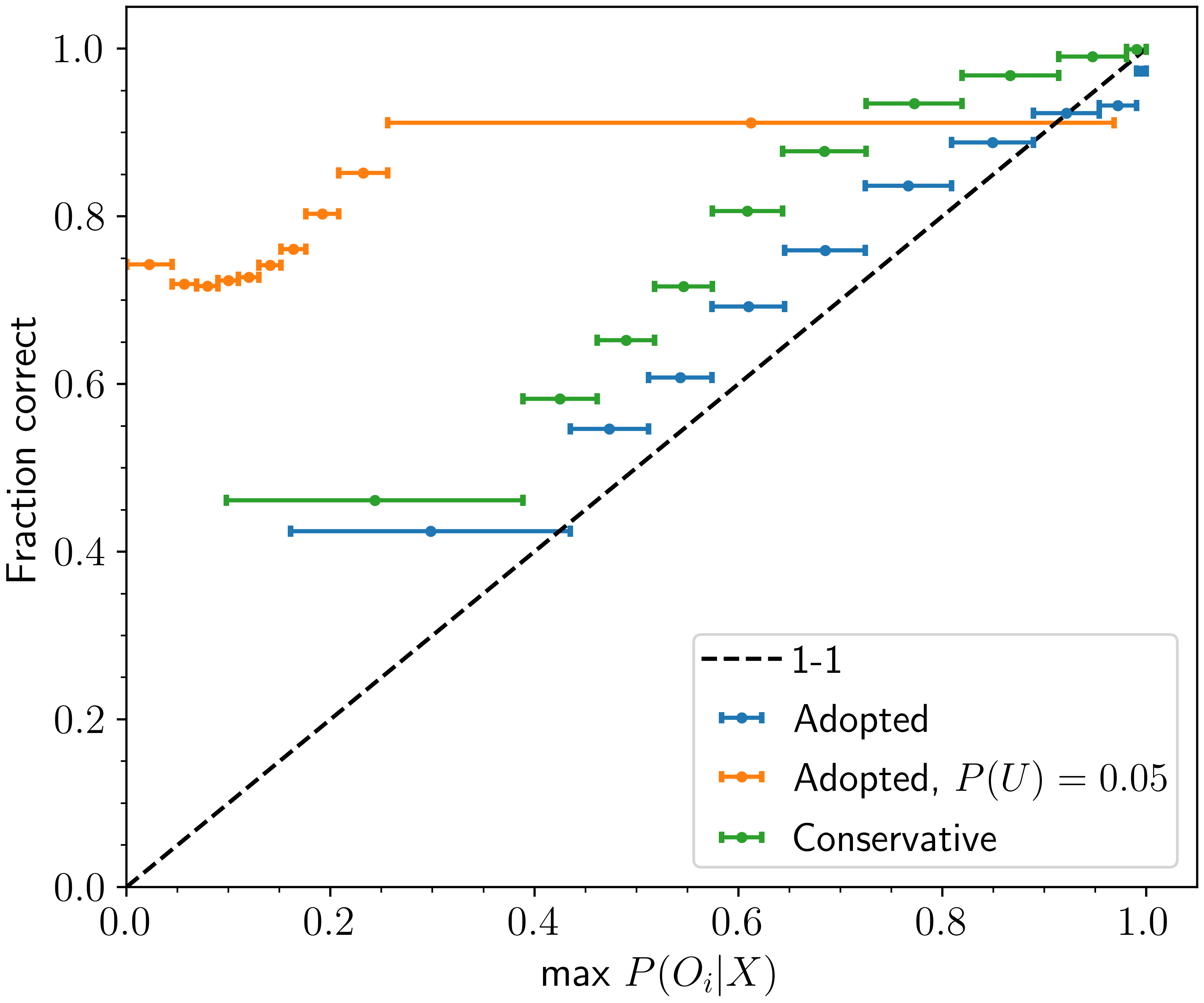}
    \caption{
    Same as Figure~\ref{fig:sb_priors}
    but for different prior assumptions as described
    in the legend.
    }
	\label{fig:sb_punchline}
\end{figure}

Figure~\ref{fig:sb_punchline} shows another set of
results but for more different choices of priors 
(Table~\ref{tab:sandbox}).
The remarkably close correspondence between the
two quantities indicates the posterior is well-calibrated,
at least for this pairing of sandbox and prior set.
We also show results for prior sets where we 
assume $\mPU = 0.05$ and for the conservative prior set
(Table~\ref{tab:priors}).
Each of these assigns systematically lower values 
to the true host galaxy yielding a higher 
percentage of correct cases at lower maximum \POix.

To characterise the behaviour of our method under different
simulated truths (i.e., sandboxes) and different priors, Table~\ref{tab:sandbox} lists
the fraction $f$ of all FRBs which are correctly identified, i.e.\ the true (T) host is securely identified
($\mPOix > 0.95$); and the fraction of secure identifications which are correct. In all cases, at least 96\% of all secure
associations find the true host, indicating that our
method is trustworthy.
The fraction of FRBs expected to have such a secure association however varies 
significantly, primarily as a function of the sandbox (variation of $\pm 0.3$ in $f(T+{\rm secure})$), with the analysis method on a given sandbox having a secondary effect (variation of $\pm 0.05$). Analysis of SB-1 demonstrates that the choice between different inverse priors has little effect, but produces a higher fraction of secure associations than a uniform prior. Interestingly, $f(T+{\rm secure})$ is much higher for the SB-4 analysis than for SB-2 and SB-3, despite all three using an assumed $p(\omega|O_i)$ of equal shape to the true $p(\omega|O_i)$, and being otherwise identical. However, while the SB-4 analysis assumes $p(\omega|O_i)$ to be uniform out to $\theta=6 \phi$, the true distribution is fully contained within $2 \phi$, unlike SB-2 and SB-3. We thus conclude that the dominant determinant of the fraction of securely (and hence, correctly) identified FRB hosts, in the case that the true host is observed, is the fraction of FRBs lying in close proximity to their hosts, irrespective of other considerations. We find it especially reassuring that results are not highly sensitive to the analysis method, i.e.\ that our formalism yields greater sensitivity to the physical truth than to our choice of reasonable priors.

\begin{figure}[!ht]
\centering
    \includegraphics[width=8.5cm]{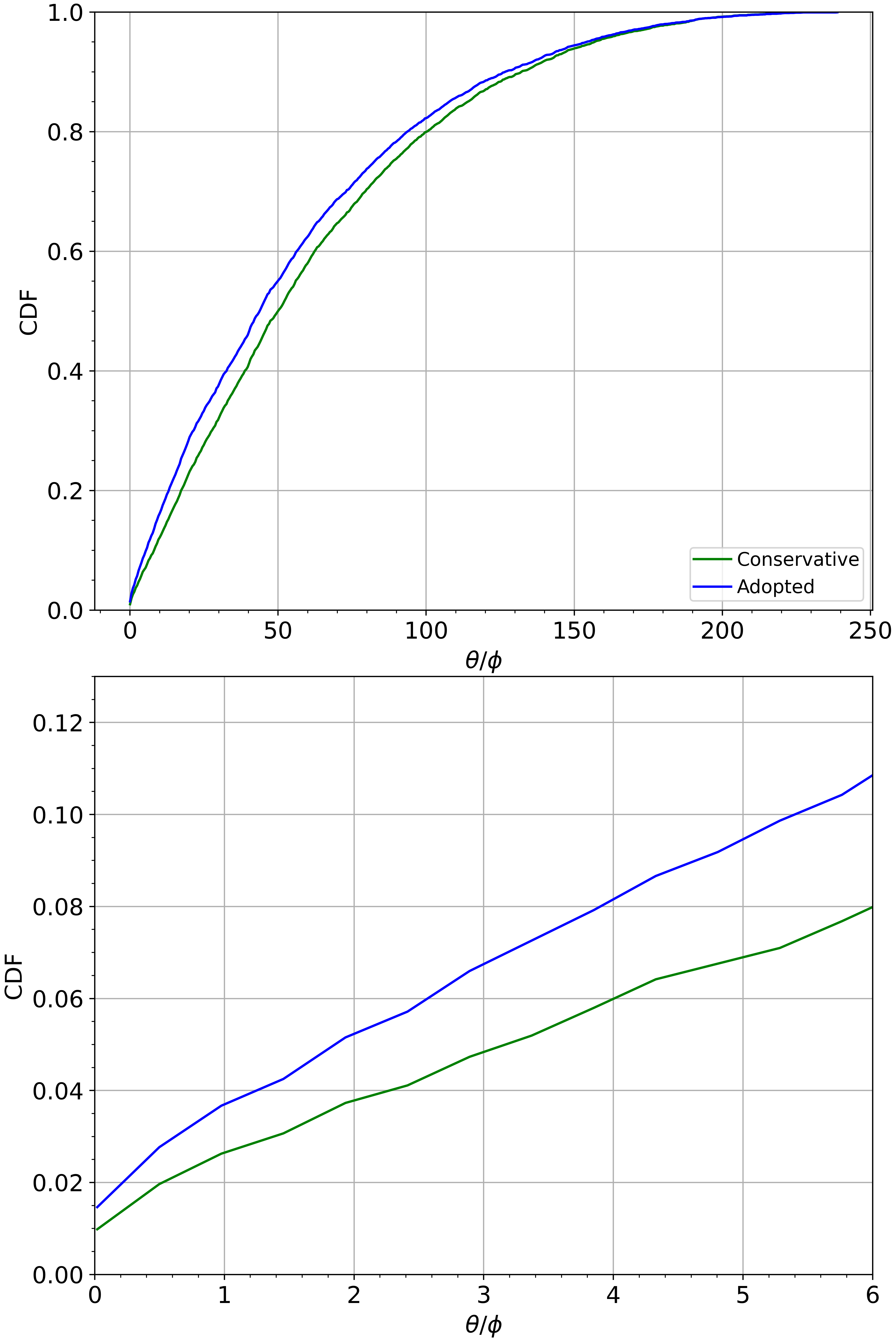}
    \caption{
    Cumulative distribution of $\theta/\phi$ for 
    candidate galaxies in SB-5 where the true host is unseen, using conservative (green) and adopted (blue) priors.
    Lower panel is a zoom-in on the region related to our adopted
    maximum offset.
    }
	\label{fig:theta_phi_unseen}
\end{figure}

What about unseen hosts? If we use $\mPU=0$, as typically assumed in this work, then $\mPUx=0$ always, and the method will tend to assign the highest posterior $\mPOix$ to the closest galaxy regardless of distance. Using the 10\% of hostless FRBs from SB-5, the conservative and adopted prior sets from Table~\ref{tab:priors} find secure associations for the majority of FRBs (55\% and 62\% respectively). However, the typical radial offset for these secure associations is very large. In Figure~\ref{fig:theta_phi_unseen}, we show the cumulative distribution of $\theta/\phi$ for such candidates. The probability of the most likely candidate being close to the FRB is small, with 10\% or less of such falsely identified hosts having $\theta/\phi < 6$. The distributions for secure associations is almost identical to that from non-secure hosts. We conclude that measuring a small $\theta/\phi$ is a strong discriminant against unseen hosts irrespective of $p(U)$.

Buoyed by these results, we now proceed to apply PATH
to real FRB observations.



%
%

\begin{deluxetable}{cccccccccccccccc}
\tablewidth{0pc}
\tablecaption{Prior Sets\label{tab:priors}}
\tabletypesize{\footnotesize}
\tablehead{\colhead{Set}  
& \colhead{\PO} & \colhead{\PU} & \colhead{\poffset}
& \colhead{\thmax/\halflight}
} 
\startdata 
Conservative&identical& 0&uniform& 6\\ 
Adopted&inverse& 0&exp& 6\\ 
\hline 
\enddata 
\end{deluxetable}

\section{Real FRB Analysis and Results}
\label{sec:results}

Informed by the results of the previous section, we 
proceed to apply the formalism to all of the published,
well-localized FRBs.  We discuss 
results for \nsets~sets of priors --- conservative and 
\adopted\  --- as summarized in Table~\ref{tab:priors}.  
We refer to the first as conservative because 
all galaxies within \thmax\ are given equal prior.

\begin{deluxetable*}{cccccccccccccccc}
\tablewidth{0pc}
\tablecaption{FRBs Analyzed\label{tab:frbs}}
\tabletypesize{\footnotesize}
\tablehead{\colhead{FRB}  
& \colhead{\rafrb} & \colhead{\decfrb}
& \colhead{\eea} & \colhead{\eeb} & \colhead{\eep}
& \colhead{Filter}
\\& (deg) & (deg) & ($''$) & ($''$) & (deg) 
} 
\startdata 
FRB121102& 82.99458& 33.14792& 0.10& 0.10& 0.0& GMOS\_N\_i \\ 
FRB180916& 29.50313& 65.71675& 0.00& 0.00& 0.0& GMOS\_N\_r \\ 
FRB180924& 326.10523& -40.90003& 0.11& 0.09& 0.0& VLT\_FORS2\_g \\ 
FRB181112& 327.34846& -52.97093& 3.25& 0.81& 120.2& VLT\_FORS2\_I \\ 
FRB190102& 322.41567& -79.47569& 0.54& 0.47& 0.0& VLT\_FORS2\_I \\ 
FRB190523& 207.06500& 72.46972& 4.00& 1.50& 340.0& LRIS\_R \\ 
FRB190608& 334.01987& -7.89825& 0.26& 0.25& 90.0& VLT\_FORS2\_I \\ 
FRB190611& 320.74546& -79.39758& 0.67& 0.67& 0.0& GMOS\_S\_i \\ 
FRB190614& 65.07552& 73.70674& 0.80& 0.40& 67.0& LRIS\_I \\ 
FRB190711& 329.41950& -80.35800& 0.40& 0.31& 90.0& GMOS\_S\_i \\ 
FRB190714& 183.97967& -13.02103& 0.36& 0.22& 90.0& VLT\_FORS2\_I \\ 
FRB191001& 323.35155& -54.74774& 0.17& 0.13& 90.0& VLT\_FORS2\_I \\ 
FRB200430& 229.70642& 12.37689& 1.07& 0.30& 0.0& LRIS\_I \\ 
\hline 
\enddata 
\tablecomments{\eea, \eeb, \eep\ define the total $1\sigma$ error ellipse for the FRB localization 
Data are taken from \cite{Ravi19,Day20,Law20,Tendulkar17,Marcote20,Heintz2020}.} 
\end{deluxetable*}

\subsection{FRB Host Candidates}

Central to the analysis is the identification and analysis
of galaxy candidates in imaging data.  The first
step --- source identification ---
is the most challenging and the most subjective.
For every image, sources near the detection limit are
subject to the precise methodology: 
background subtraction, thresholding, pixel grouping, and deblending.
After experimenting with the routines encoded in the
{\sc photutils} package, we settled on the following key parameters: \texttt{npixels} $=9$, \texttt{deblend}$=$True, \texttt{xy\_kernel} $=(3,3)$,
\texttt{Gaussian2Dkernel}, \texttt{nsig} $=3.$ (kernel),  \texttt{nsig} $=1.5$ (threshold),
\texttt{background} $=(50,50)$, \texttt{filter\_size} $=(3,3)$, median background.

To test these choices, we independently analyzed the
data with the SExtractor package using a standard set
of input parameters. Namely, we set {\texttt {DETECT\_MINAREA}} $= 9$ and {\texttt {DETECT\_THRES}} $= 1.5$ for consistency with the {\sc photutils} parameters. Images are filtered with the default convolution kernel (\texttt{default.conv}). To recover blended sources (see e.g., FRB 180924 below), we set {\texttt {DEBLEND\_MINCONT}} $= 0.0001$.

With this set of parameters, we recover the centroid positions within $\approx 2\%$, while the aperture sizes show scatter up to $\approx 15\%$. We note that slight differences between the {\sc photutils} and SExtractor methodologies may be driving these discrepancies; namely, while SExtractor uses a multi-thresholding deblending technique, {\sc photutils} utilizes a combination of multi-thresholding and watershed segmentation. Furthermore, the \texttt{default.conv} convolution kernel is equivalent to a $3\times3$ Gaussian kernel with FWHM $ = 2$, in slight contrast to the Gaussian kernel used above. Nevertheless, we find comparable results using the two methods.

\begin{figure}[!ht]
\centering
    \includegraphics[width=8.5cm]{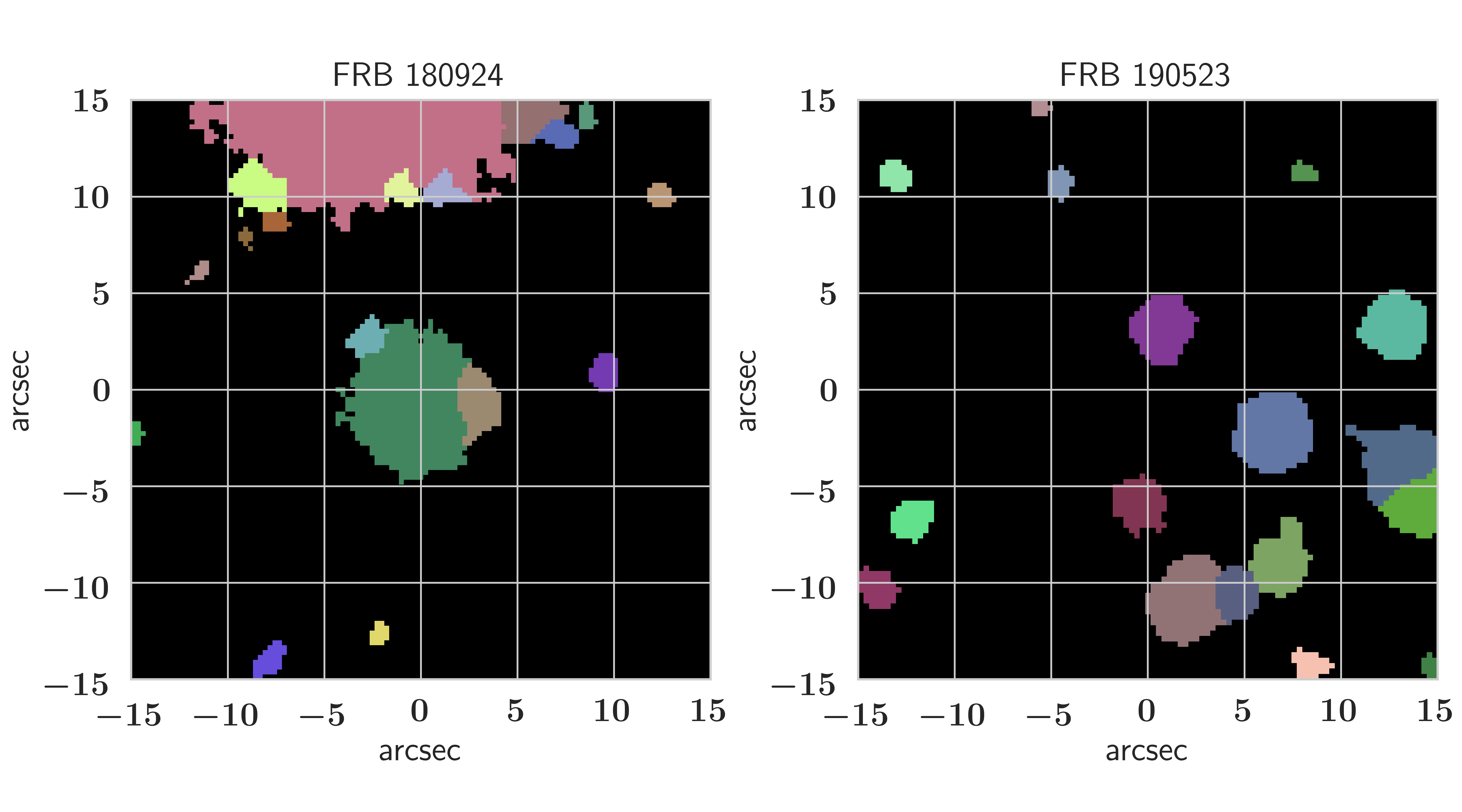}
    \caption{
    Segmentation images for the sources in $30''$
    cutouts around FRB~180924 (left) and FRB~190523 (right).
    The multitude of sources (several erroneous) at the 
    top of the FRB180924 image is due to artefacts from a very
    bright star.
    }
	\label{fig:segm}
\end{figure}

Figure~\ref{fig:segm} shows the segmentation maps of
FRB~180924 and FRB~190523.
Note the three, blended sources near the localization of
FRB~180924 which are known to be unique galaxies
at distinct redshifts \citep{Bannister19}.  An image
with shallower depth (e.g.\ DES-DR1) or a different choice
of {\sc photutils} parameters would lead to the non-detection
of the fainter sources. 
This highlights the subjectivity of
source identification that can affect the final results.

The source identification packages offer an assessment of
the source shape (e.g.\ ellipticity and size) which can be used to 
used to select and then ignore Galactic stars.   
For the analysis that follows, we have simply
clipped bright stars according to their apparent
magnitudes when necessary.

Provided with the segmentation map, one may perform aperture
photometry and estimate \halflight\ from the derived
elliptical apertures.  All of the measurements for the 
galaxy candidates are provided
in Table~\ref{tab:results}.

\begin{figure*}[!ht]
\centering
    \includegraphics[width=\textwidth]{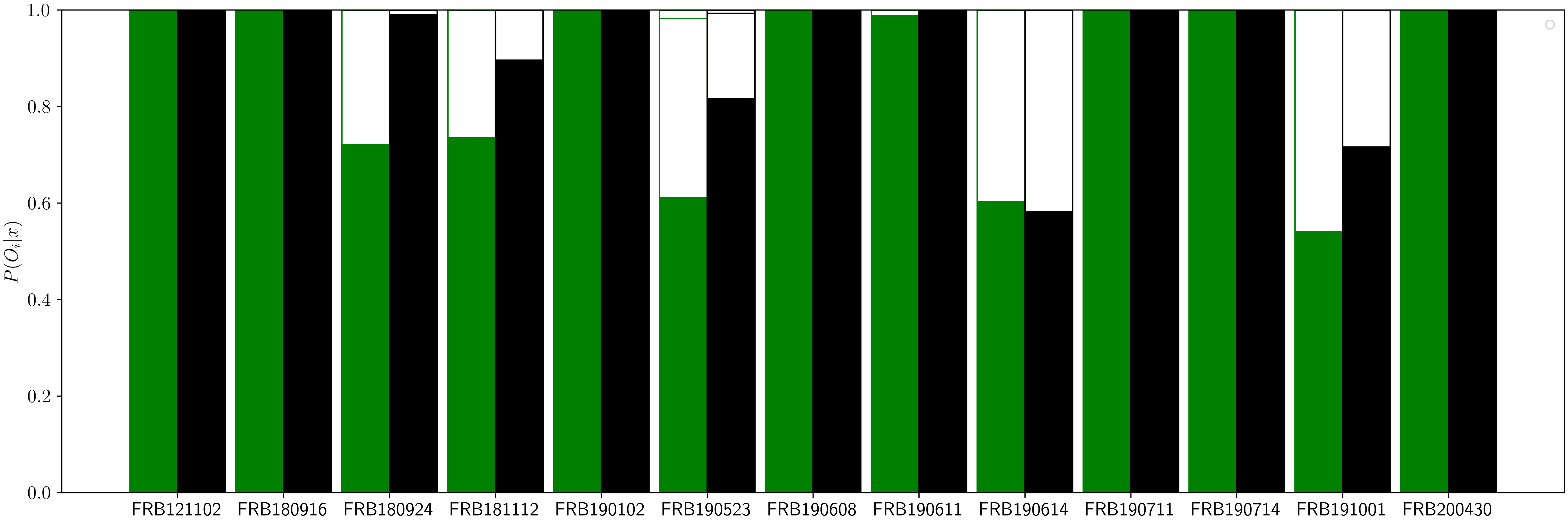}
    \caption{
    Posterior probabilities \POix\ for the most likely host (solid bars) and all other
    candidates (open bars) with $\mPOix > \mPOsec = \POvsec$,
    as a function of prior set
    (green = conservative; black=adopted).
    With $\mPOsec = \POvsec$ as the probability for a 
    secure association, there are currently \nsecure\ FRBs 
    satisfying this criterion using the adopted prior set.  
    The non-secure hosts occur for a variety of reasons as
    described in the text.
    }
	\label{fig:POix}
\end{figure*}

\begin{figure*}[!ht]
\centering
    \includegraphics[width=\textwidth]{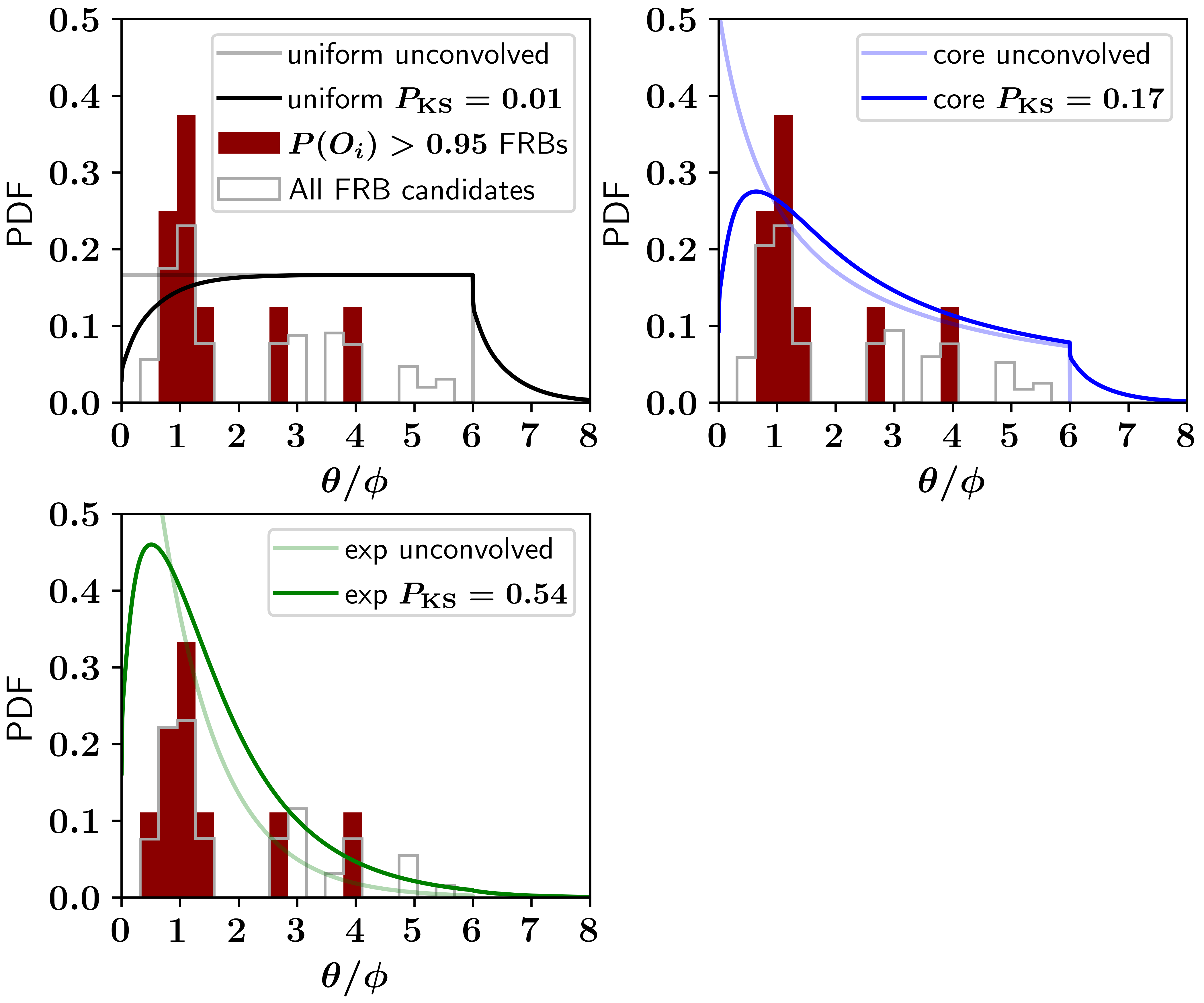}
    \caption{
    The solid histogram shows the distribution of separations for the secure
    host galaxies, in units of their angular size (\halflight).
    The gray histogram is for all of the candidate galaxies but weighted
    by their posterior probabilities \POix.
    These results were derived by assuming the 
    \adopted\ prior set (Table~\ref{tab:priors}) and by varying the 
    offset function \poffset, as labeled in each panel. 
    Overplotted on the histogram is the offset function both before
    (semi-transparent) and after convolving with the FRB localization 
    error (solid).  The data rule out 
    the uniform offset function that extends to
    $\mthmax = 6 \phi$  at $\approx 99\%$c.l.
    (using a one-sided Kolmogorov-Smirnov test). 
    } 
	\label{fig:poff_posterior}
\end{figure*}

\subsection{FRB Assignments}

We then applied the PATH framework. 
The primary results are displayed in Figure~\ref{fig:POix}
which summarizes the \POix\ values for the candidates
in each field for each set of priors.
We find very similar results for the \nsets~prior sets
with one obvious exception: FRB~180924.
In this case, there are two galaxies with separation
$\theta < \mthmax$ 
from this precisely localized FRB.
These are treated similarly by the conservative approach.
We demonstrate below, however, that a uniform \poffset\ 
function with $\mthmax = 6 \phi$
is disfavored by the data.
Imposing the exponential offset model yields 
a higher \POix\ for the primary candidate.
Furthermore, if we allow for the great difference
in apparent magnitude by invoking the 
inverse \PO\ prior, the posterior
\POix\ raises to near unity for the host
reported by \cite{Bannister19}.

Based on the results from analysis of mock fields
($\S$~\ref{sec:sim}), we adopt a probability 
threshold $\mPOsec = \POvsec$ above which we 
consider a host association to be highly secure.
The results in Figure~\ref{fig:POix}
indicate that \nsecure\ of the FRBs are
associated to a single galaxy with $\mPOix > \POvsec$ 
for the \adopted\ prior set 
(and eight for the conservative set with FRB~180924 the 
difference).

The non-secure hosts
deserve individual consideration,
in part to understand the dependence of the formalism
to observed variations on the sky.
FRB~181112 shows two bright galaxies near the FRB with the
brighter assumed to lie in the foreground
\citep{Prochaska19b}.  
The PATH analysis for 
the purported host gives $\mPOix \approx 0.7-0.9$
depending on the choice of priors (Table~\ref{tab:priors}).
We emphasize that the majority of FRB sightlines that
intersect a massive foreground halo (estimated to 
be a few percent for FRBs at $z>0.5$), will tend to 
have a maximum $\mPOix < \mPOsec$.  
Given the terrific scientific value of probing such
halos with FRBs \citep{Prochaska19b}, one may need
to introduce additional criteria/priors to confidently
pursue this science.

The next, non-secure FRB association is FRB~190523
whose larger localization error incorporates several
candidates. 
The analysis, however, does favor the purported host
reported by \cite{Ravi19}.
Third is FRB~190614 which lies near ($\lesssim 2''$) two 
faint galaxies with unknown redshifts \citep{Law20}.
As the FRB with the highest DM and therefore the highest
presumed redshift of the sample, this result emphasizes
the likely challenges of associating high-$z$ FRBs
to galaxies.  In particular, given the host itself is
likely faint, the incidence of additional, chance
associations with comparable \POix\ will be higher.
Last is FRB~191001 which sits next to two bright galaxies
known to have a common redshift \citep{Bhandari20b}.
Therefore, 
the redshift of the FRB is secure but the
host offset and its internal properties (e.g.\ stellar
mass) are currently based on the assumption that the
closer galaxy is the host 
and, indeed, it exhibits a $3\times$ higher \POix\ value
for the \adopted\ prior set.


\subsection{Towards Additional Priors}
\label{sec:add_priors}

Having established a set of \nsecure\ secure, 
$\mPOix > \mPOsec = \POvsec$, 
host associations we may 
test the assumed \poffset\ functions 
imposed in the analysis.
Figure~\ref{fig:poff_posterior} shows the offset distribution
for the secure hosts for the three \poffset\ priors
of the analysis ($\S$~\ref{sec:poffset}).
Note that modifying the choice of \poffset\ could include/exclude
FRBs as being secure.  The figure also shows the 
values of $\theta/\mhalflight$ derived for all candidates
from the full set of 
FRBs, where we have weighted the $\theta/\mhalflight$ value
of each candidate by $\mPOix$.  
Overall, the posteriors lend reasonable credibility to the 
set of \poffset\ functions.  On the other hand, a comparison
of the secure distribution with the priors yields 
a one-sided Kolmogorov-Smirnov probability 
$P_{\rm KS} \lesssim 0.1$ and we rule out the uniform prior
to $\mthmax = 6 \phi$ at $> 99\%$.  
The data appear to favor a \poffset\ function that favors
a central concentration for FRB locations.
Additionally, such small values of $\theta/\mhalflight$ are unlikely when the true host galaxy is unseen (see Figure~\ref{fig:theta_phi_unseen}), being $<3\%$ for the seven secure hosts with $\theta/\mhalflight < 1.5$. Since all FRBs have most likely candidates with $\theta/\mhalflight < 5$, we conclude that no more than one of the FRBs considered can have an unseen host ($p \lesssim 0.01$).

\begin{figure}[!ht]
\centering
    \includegraphics[width=0.48\textwidth]{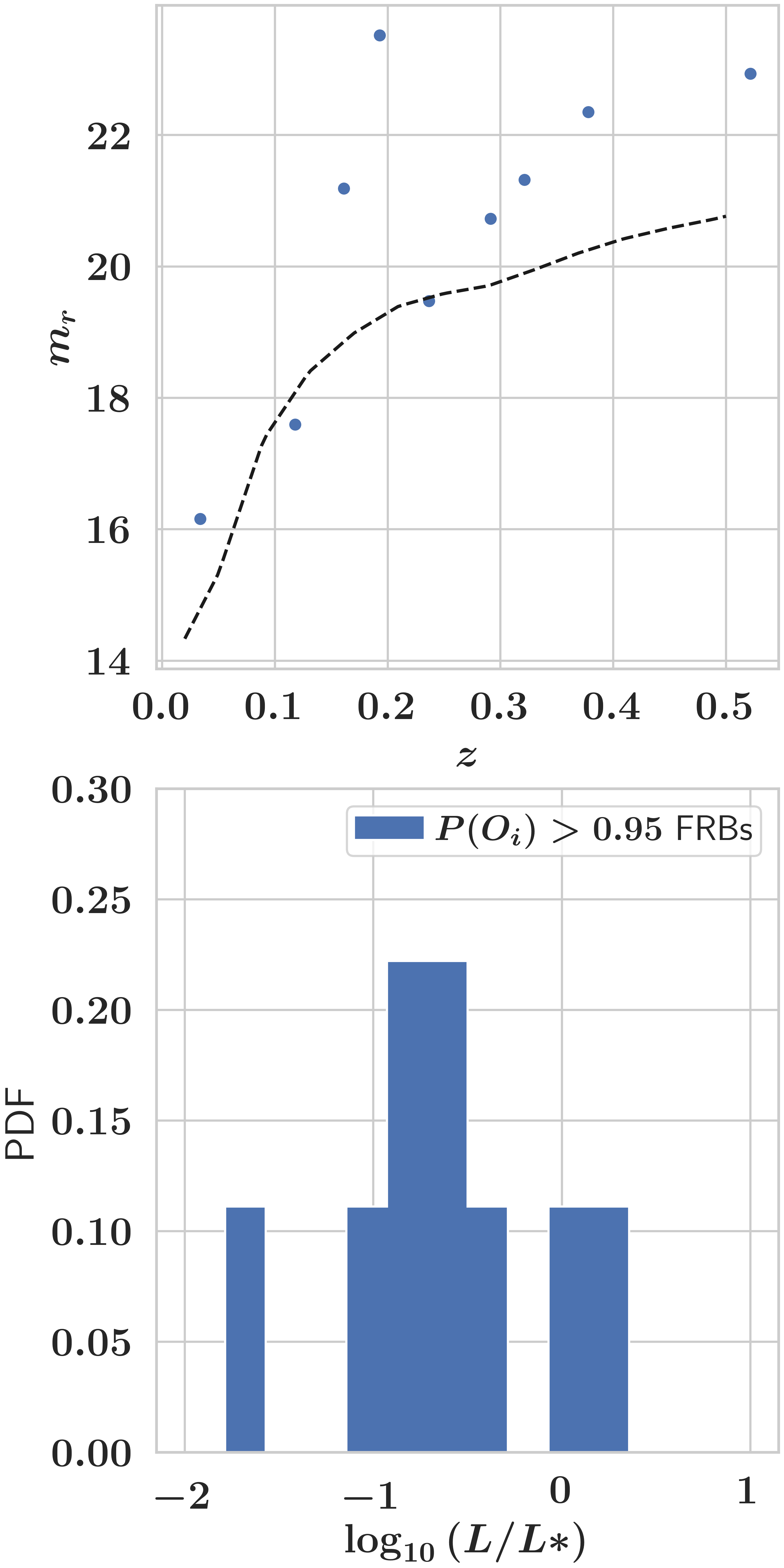}
    \caption{
    ({\it top}) Scatter plot of apparent magnitudes versus redshift
    for the \nsecure\ host galaxies.  These show an expected
    decrease in flux with increasing distance.
    The dashed line marks the approximate apparent
    magnitude for an $L*$ galaxy.
    ({\it bottom}) Estimated galaxy luminosity relative to the
    characteristic luminosity $L*$ at the host redshift.
    The secure hosts have a median $L/L* \approx 1/4$ and
    an RMS scatter of 0.5\,dex.
    }
	\label{fig:PL}
\end{figure}

Encoded in every FRB is its dispersion measure (DM),
the path integral of free electrons along the sightline
weighted by the cosmological scale factor.  The first
$\approx 10$ FRBs localized have established a 
firm correlation between DM and redshift, 
now termed the Macquart Relation \citep{Macquart20}.
This relation derives from the ionized plasma
that permeates the cosmic web.
Because redshift (i.e.\ distance)
affects observed properties, one
should consider incorporating the DM
into the association analysis. 
A full and proper treatment, however, requires including
the intrinsic luminosities and spectral slopes of FRBs 
convolved with instrumental sensitivity and even the
triggering software (James et al., in prep.).
Further, we emphasize that adopting the Macquart Relation
as a prior would likely require estimating the redshift
for every galaxy candidate;  this will be intractable
for many FRBs.

\begin{figure*}[!ht]
\centering
    \includegraphics[width=\textwidth]{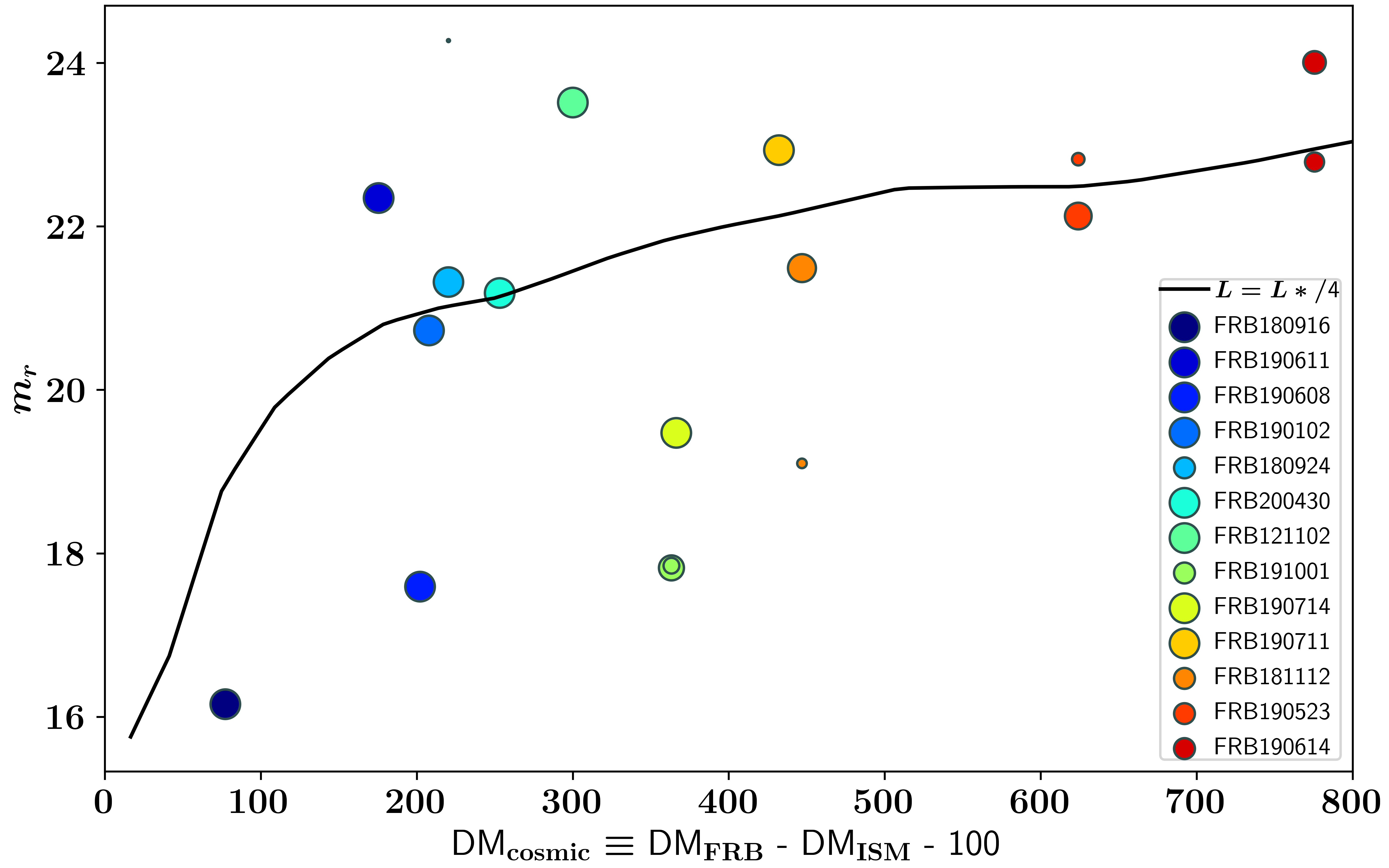}
    \caption{
    Using the adopted prior set (Table~\ref{tab:priors}), 
    we show the apparent magnitudes of each candidate
    with $\mPOix > 0.01$ scaling the point-size by \POix.  These are plotted
    against \dmcosmic, an estimate of the dispersion measure of the FRB
    due to the cosmic web (see text for details).
    The black curve shows the estimated $m_r$ for an
    $L = L*/4$ galaxy by converting \dmcosmic\ to redshift
    (using the Macquart relation).  
    The fact that the $m_r$ values track this curve
    implies an underlying Macquart relation.
    }
	\label{fig:mag_DM}
\end{figure*}

In lieu of a direct application of the Macquart relation,
we propose to leverage the luminosity distribution
(Figure~\ref{fig:PL}) together with DM.
Figure~\ref{fig:mag_DM} presents the apparent magnitudes of 
each galaxy candidate against an estimate of the cosmic 
dispersion measure \dmcosmic\ with the 
point-size proportional to \POix.
For \dmcosmic, we adopt a simple estimation: 

\begin{equation}
    \mdmcosmic = \mdmfrb - \mdmmwism - 100 
\end{equation}
with  \dmmwism\ the estimated ISM dispersion measure \citep{ne2001}
and the factor of 100 DM units accounts for the Galactic
halo and the host galaxy \citep[see][]{xyz19,Macquart20}.
The locus of data exhibits a clear correlation reflecting the
decrease in observed galaxy flux with increasing distance.

Converting the \dmcosmic\ estimates to 
redshift\footnote{https://github.com/FRBs/FRB/blob/main/docs/nb/DM\_cosmic.ipynb},
we may convert a given $L_r$ luminosity to $m_r$;
this is illustrated as the black curve in Figure~\ref{fig:mag_DM}
where we assumed a fiducial $L = L*/4$
based on Figure~\ref{fig:PL}.
The good correspondence between the data and this curve
reveals the Macquart relation without having used 
any direct redshift measurements.

In principle, one could construct a prior 
$P(m_r|{\rm DM})$ to include in the analysis.
This will, however, be subject to scatter in 
\dmcosmic\ \citep{Macquart20}, \dmhost, 
and the intrinsic luminosities of the host galaxy
population (Figure~\ref{fig:PL}).
It will also be subject, however, to the S/N
considerations that affect any prior related to DM
\citep{James20}.

\startlongtable
\begin{deluxetable*}{cccccccccccccccc}
\tablewidth{0pc}
\tablecaption{Results for FRB Associations\label{tab:results}}
\tabletypesize{\footnotesize}
\tablehead{\colhead{FRB}  
& \colhead{RA$_{\rm cand}$} & \colhead{Dec$_{\rm cand}$} 
& \colhead{$\theta$} 
& \colhead{\halflight}  
& \colhead{\gmag} 
& \colhead{Filter} & \colhead{\pchance} 
& \colhead{\PO} & \colhead{\POx} 
& \colhead{\PU} & \colhead{\PUx} 
\\} 
\startdata 
\cutinhead{Conservative} 
FRB121102& 82.9945 & $33.1479$& 0.2& 0.28& 23.52& GMOS\_N\_i& 0.0039& 0.1000& 1.0000& 0.0000& 0.0000\\ 
& 82.9942 & $33.1472$& 2.9& 0.28& 21.14& GMOS\_N\_i& 0.0113& 0.1000& 0.0000& 0.0000& 0.0000\\ 
& 82.9935 & $33.1473$& 3.9& 0.23& 24.18& GMOS\_N\_i& 0.2487& 0.1000& 0.0000& 0.0000& 0.0000\\ 
& 82.9960 & $33.1485$& 4.7& 0.25& 23.28& GMOS\_N\_i& 0.1818& 0.1000& 0.0000& 0.0000& 0.0000\\ 
& 82.9939 & $33.1492$& 4.8& 0.13& 25.06& GMOS\_N\_i& 0.5740& 0.1000& 0.0000& 0.0000& 0.0000\\ 
& 82.9923 & $33.1469$& 7.9& 0.28& 21.58& GMOS\_N\_i& 0.1169& 0.1000& 0.0000& 0.0000& 0.0000\\ 
& 82.9968 & $33.1490$& 7.7& 0.24& 22.91& GMOS\_N\_i& 0.3195& 0.1000& 0.0000& 0.0000& 0.0000\\ 
& 82.9948 & $33.1503$& 8.5& 0.15& 24.95& GMOS\_N\_i& 0.9101& 0.1000& 0.0000& 0.0000& 0.0000\\ 
& 82.9918 & $33.1470$& 9.2& 0.25& 23.53& GMOS\_N\_i& 0.6051& 0.1000& 0.0000& 0.0000& 0.0000\\ 
& 82.9937 & $33.1453$& 10.0& 0.31& 20.19& GMOS\_N\_i& 0.0501& 0.1000& 0.0000& 0.0000& 0.0000\\ 
FRB180916& 29.5012 & $65.7148$& 7.7& 3.03& 16.16& GMOS\_N\_r& 0.0005& 0.0588& 1.0000& 0.0000& 0.0000\\ 
& 29.5054 & $65.7140$& 10.5& 0.53& 21.42& GMOS\_N\_r& 0.1728& 0.0588& 0.0000& 0.0000& 0.0000\\ 
& 29.5093 & $65.7179$& 10.0& 0.21& 22.00& GMOS\_N\_r& 0.2554& 0.0588& 0.0000& 0.0000& 0.0000\\ 
& 29.4998 & $65.7130$& 14.4& 0.66& 20.96& GMOS\_N\_r& 0.2068& 0.0588& 0.0000& 0.0000& 0.0000\\ 
& 29.5084 & $65.7139$& 12.7& 0.24& 22.71& GMOS\_N\_r& 0.5888& 0.0588& 0.0000& 0.0000& 0.0000\\ 
& 29.4913 & $65.7174$& 17.6& 1.02& 20.91& GMOS\_N\_r& 0.2816& 0.0588& 0.0000& 0.0000& 0.0000\\ 
& 29.5060 & $65.7211$& 16.3& 0.41& 21.29& GMOS\_N\_r& 0.3312& 0.0588& 0.0000& 0.0000& 0.0000\\ 
& 29.4996 & $65.7203$& 13.7& 0.39& 19.83& GMOS\_N\_r& 0.0656& 0.0588& 0.0000& 0.0000& 0.0000\\ 
& 29.5129 & $65.7192$& 17.1& 0.37& 20.03& GMOS\_N\_r& 0.1202& 0.0588& 0.0000& 0.0000& 0.0000\\ 
& 29.5136 & $65.7181$& 16.3& 0.42& 19.42& GMOS\_N\_r& 0.0598& 0.0588& 0.0000& 0.0000& 0.0000\\ 
& 29.5052 & $65.7125$& 15.7& 0.43& 18.15& GMOS\_N\_r& 0.0141& 0.0588& 0.0000& 0.0000& 0.0000\\ 
& 29.5122 & $65.7171$& 13.5& 0.47& 19.02& GMOS\_N\_r& 0.0274& 0.0588& 0.0000& 0.0000& 0.0000\\ 
& 29.5015 & $65.7124$& 15.7& 0.16& 21.13& GMOS\_N\_r& 0.2748& 0.0588& 0.0000& 0.0000& 0.0000\\ 
& 29.4933 & $65.7138$& 18.0& 0.33& 20.19& GMOS\_N\_r& 0.1541& 0.0588& 0.0000& 0.0000& 0.0000\\ 
& 29.5110 & $65.7136$& 16.4& 0.19& 22.17& GMOS\_N\_r& 0.6012& 0.0588& 0.0000& 0.0000& 0.0000\\ 
& 29.5021 & $65.7130$& 13.6& 0.59& 21.73& GMOS\_N\_r& 0.3480& 0.0588& 0.0000& 0.0000& 0.0000\\ 
& 29.5029 & $65.7218$& 18.3& 0.28& 21.64& GMOS\_N\_r& 0.5053& 0.0588& 0.0000& 0.0000& 0.0000\\ 
FRB180924& 326.1042 & $-40.9002$& 2.9& 0.81& 24.27& VLT\_FORS2\_g& 0.1973& 0.2500& 0.7172& 0.0000& 0.0000\\ 
& 326.1054 & $-40.9002$& 0.8& 1.31& 21.32& VLT\_FORS2\_g& 0.0118& 0.2500& 0.2779& 0.0000& 0.0000\\ 
& 326.1062 & $-40.8993$& 3.8& 0.50& 25.47& VLT\_FORS2\_g& 0.5390& 0.2500& 0.0049& 0.0000& 0.0000\\ 
& 326.1017 & $-40.8998$& 9.5& 0.46& 25.30& VLT\_FORS2\_g& 0.9807& 0.2500& 0.0000& 0.0000& 0.0000\\ 
FRB181112& 327.3486 & $-52.9709$& 0.4& 0.67& 21.49& VLT\_FORS2\_I& 0.0227& 0.2500& 0.7588& 0.0000& 0.0000\\ 
& 327.3496 & $-52.9696$& 5.4& 1.06& 19.10& VLT\_FORS2\_I& 0.0073& 0.2500& 0.2411& 0.0000& 0.0000\\ 
& 327.3484 & $-52.9729$& 7.0& 0.58& 22.01& VLT\_FORS2\_I& 0.1646& 0.2500& 0.0001& 0.0000& 0.0000\\ 
& 327.3467 & $-52.9727$& 7.4& 0.32& 24.05& VLT\_FORS2\_I& 0.6612& 0.2500& 0.0000& 0.0000& 0.0000\\ 
FRB190102& 322.4149 & $-79.4756$& 0.5& 0.86& 20.73& VLT\_FORS2\_I& 0.0038& 0.5000& 1.0000& 0.0000& 0.0000\\ 
& 322.4173 & $-79.4773$& 5.9& 0.55& 22.54& VLT\_FORS2\_I& 0.1623& 0.5000& 0.0000& 0.0000& 0.0000\\ 
FRB190523& 207.0642 & $72.4706$& 3.4& 0.71& 22.13& LRIS\_R& 0.1158& 0.3333& 0.6116& 0.0000& 0.0000\\ 
& 207.0654 & $72.4681$& 5.8& 0.61& 22.82& LRIS\_R& 0.2986& 0.3333& 0.3712& 0.0000& 0.0000\\ 
& 207.0589 & $72.4691$& 6.9& 0.72& 20.78& LRIS\_R& 0.0664& 0.3333& 0.0173& 0.0000& 0.0000\\ 
FRB190608& 334.0203 & $-7.8988$& 2.5& 1.66& 17.60& VLT\_FORS2\_I& 0.0005& 0.2500& 1.0000& 0.0000& 0.0000\\ 
& 334.0185 & $-7.8986$& 5.0& 0.30& 24.83& VLT\_FORS2\_I& 0.5373& 0.2500& 0.0000& 0.0000& 0.0000\\ 
& 334.0186 & $-7.8969$& 6.6& 0.26& 25.28& VLT\_FORS2\_I& 0.8461& 0.2500& 0.0000& 0.0000& 0.0000\\ 
& 334.0187 & $-7.8959$& 9.5& 0.45& 22.76& VLT\_FORS2\_I& 0.4081& 0.2500& 0.0000& 0.0000& 0.0000\\ 
FRB190611& 320.7429 & $-79.3973$& 2.0& 0.50& 22.35& GMOS\_S\_i& 0.0267& 0.0909& 0.9480& 0.0000& 0.0000\\ 
& 320.7495 & $-79.3972$& 3.1& 0.27& 25.87& GMOS\_S\_i& 0.5297& 0.0909& 0.0359& 0.0000& 0.0000\\ 
& 320.7439 & $-79.3985$& 3.3& 0.26& 24.91& GMOS\_S\_i& 0.3446& 0.0909& 0.0140& 0.0000& 0.0000\\ 
& 320.7539 & $-79.3979$& 5.7& 0.65& 23.63& GMOS\_S\_i& 0.3518& 0.0909& 0.0019& 0.0000& 0.0000\\ 
& 320.7383 & $-79.3977$& 4.8& 0.36& 23.44& GMOS\_S\_i& 0.2267& 0.0909& 0.0001& 0.0000& 0.0000\\ 
& 320.7346 & $-79.3988$& 8.4& 0.53& 23.36& GMOS\_S\_i& 0.5033& 0.0909& 0.0000& 0.0000& 0.0000\\ 
& 320.7541 & $-79.3965$& 7.0& 0.18& 26.52& GMOS\_S\_i& 0.9936& 0.0909& 0.0000& 0.0000& 0.0000\\ 
& 320.7569 & $-79.3991$& 9.4& 0.52& 23.65& GMOS\_S\_i& 0.6643& 0.0909& 0.0000& 0.0000& 0.0000\\ 
& 320.7364 & $-79.3998$& 9.9& 0.41& 24.56& GMOS\_S\_i& 0.9166& 0.0909& 0.0000& 0.0000& 0.0000\\ 
& 320.7319 & $-79.3970$& 9.2& 0.27& 25.04& GMOS\_S\_i& 0.9551& 0.0909& 0.0000& 0.0000& 0.0000\\ 
& 320.7587 & $-79.3986$& 9.5& 0.17& 26.91& GMOS\_S\_i& 1.0000& 0.0909& 0.0000& 0.0000& 0.0000\\ 
FRB190614& 65.0743 & $73.7068$& 1.3& 0.41& 24.01& LRIS\_I& 0.0552& 0.2000& 0.6032& 0.0000& 0.0000\\ 
& 65.0738 & $73.7064$& 2.2& 0.40& 22.79& LRIS\_I& 0.0386& 0.2000& 0.3968& 0.0000& 0.0000\\ 
& 65.0705 & $73.7075$& 5.7& 0.33& 24.26& LRIS\_I& 0.4949& 0.2000& 0.0000& 0.0000& 0.0000\\ 
& 65.0817 & $73.7079$& 7.5& 0.17& 26.35& LRIS\_I& 0.9945& 0.2000& 0.0000& 0.0000& 0.0000\\ 
& 65.0691 & $73.7081$& 8.2& 0.28& 25.21& LRIS\_I& 0.9387& 0.2000& 0.0000& 0.0000& 0.0000\\ 
FRB190711& 329.4194 & $-80.3581$& 0.5& 0.46& 22.93& GMOS\_S\_i& 0.0108& 0.2000& 0.8821& 0.0000& 0.0000\\ 
& 329.4187 & $-80.3586$& 2.1& 0.26& 24.88& GMOS\_S\_i& 0.1471& 0.2000& 0.1179& 0.0000& 0.0000\\ 
& 329.4143 & $-80.3570$& 4.7& 0.22& 24.69& GMOS\_S\_i& 0.4654& 0.2000& 0.0000& 0.0000& 0.0000\\ 
& 329.4117 & $-80.3571$& 5.7& 0.29& 24.88& GMOS\_S\_i& 0.6545& 0.2000& 0.0000& 0.0000& 0.0000\\ 
& 329.4190 & $-80.3595$& 5.3& 0.25& 23.97& GMOS\_S\_i& 0.3682& 0.2000& 0.0000& 0.0000& 0.0000\\ 
FRB190714& 183.9795 & $-13.0212$& 1.0& 0.95& 19.48& VLT\_FORS2\_I& 0.0012& 0.2000& 1.0000& 0.0000& 0.0000\\ 
& 183.9797 & $-13.0193$& 6.1& 0.54& 23.71& VLT\_FORS2\_I& 0.3925& 0.2000& 0.0000& 0.0000& 0.0000\\ 
& 183.9787 & $-13.0229$& 7.4& 0.60& 21.22& VLT\_FORS2\_I& 0.0772& 0.2000& 0.0000& 0.0000& 0.0000\\ 
& 183.9797 & $-13.0230$& 7.0& 0.31& 24.36& VLT\_FORS2\_I& 0.6494& 0.2000& 0.0000& 0.0000& 0.0000\\ 
& 183.9795 & $-13.0234$& 8.7& 0.50& 22.71& VLT\_FORS2\_I& 0.3425& 0.2000& 0.0000& 0.0000& 0.0000\\ 
FRB191001& 323.3525 & $-54.7487$& 3.9& 1.36& 17.82& VLT\_FORS2\_I& 0.0009& 0.3333& 0.5412& 0.0000& 0.0000\\ 
& 323.3492 & $-54.7483$& 5.3& 1.47& 17.85& VLT\_FORS2\_I& 0.0015& 0.3333& 0.4588& 0.0000& 0.0000\\ 
& 323.3501 & $-54.7496$& 7.3& 0.27& 25.11& VLT\_FORS2\_I& 0.8690& 0.3333& 0.0000& 0.0000& 0.0000\\ 
FRB200430& 229.7064 & $12.3766$& 0.9& 0.72& 21.19& LRIS\_I& 0.0056& 0.5000& 1.0000& 0.0000& 0.0000\\ 
& 229.7088 & $12.3778$& 8.9& 0.38& 24.82& LRIS\_I& 0.9123& 0.5000& 0.0000& 0.0000& 0.0000\\ 
\cutinhead{Adopted} 
FRB121102& 82.9945 & $33.1479$& 0.2& 0.28& 23.52& GMOS\_N\_i& 0.0039& 0.0245& 1.0000& 0.0000& 0.0000\\ 
& 82.9942 & $33.1472$& 2.9& 0.28& 21.14& GMOS\_N\_i& 0.0113& 0.2026& 0.0000& 0.0000& 0.0000\\ 
& 82.9935 & $33.1473$& 3.9& 0.23& 24.18& GMOS\_N\_i& 0.2487& 0.0144& 0.0000& 0.0000& 0.0000\\ 
& 82.9960 & $33.1485$& 4.7& 0.25& 23.28& GMOS\_N\_i& 0.1818& 0.0298& 0.0000& 0.0000& 0.0000\\ 
& 82.9939 & $33.1492$& 4.8& 0.13& 25.06& GMOS\_N\_i& 0.5740& 0.0073& 0.0000& 0.0000& 0.0000\\ 
& 82.9923 & $33.1469$& 7.9& 0.28& 21.58& GMOS\_N\_i& 0.1169& 0.1332& 0.0000& 0.0000& 0.0000\\ 
& 82.9968 & $33.1490$& 7.7& 0.24& 22.91& GMOS\_N\_i& 0.3195& 0.0408& 0.0000& 0.0000& 0.0000\\ 
& 82.9948 & $33.1503$& 8.5& 0.15& 24.95& GMOS\_N\_i& 0.9101& 0.0079& 0.0000& 0.0000& 0.0000\\ 
& 82.9918 & $33.1470$& 9.2& 0.25& 23.53& GMOS\_N\_i& 0.6051& 0.0242& 0.0000& 0.0000& 0.0000\\ 
& 82.9937 & $33.1453$& 10.0& 0.31& 20.19& GMOS\_N\_i& 0.0501& 0.5152& 0.0000& 0.0000& 0.0000\\ 
FRB180916& 29.5012 & $65.7148$& 7.7& 3.03& 16.16& GMOS\_N\_r& 0.0005& 0.8200& 1.0000& 0.0000& 0.0000\\ 
& 29.5054 & $65.7140$& 10.5& 0.53& 21.42& GMOS\_N\_r& 0.1728& 0.0026& 0.0000& 0.0000& 0.0000\\ 
& 29.5093 & $65.7179$& 10.0& 0.21& 22.00& GMOS\_N\_r& 0.2554& 0.0015& 0.0000& 0.0000& 0.0000\\ 
& 29.4998 & $65.7130$& 14.4& 0.66& 20.96& GMOS\_N\_r& 0.2068& 0.0040& 0.0000& 0.0000& 0.0000\\ 
& 29.5084 & $65.7139$& 12.7& 0.24& 22.71& GMOS\_N\_r& 0.5888& 0.0008& 0.0000& 0.0000& 0.0000\\ 
& 29.4913 & $65.7174$& 17.6& 1.02& 20.91& GMOS\_N\_r& 0.2816& 0.0042& 0.0000& 0.0000& 0.0000\\ 
& 29.5060 & $65.7211$& 16.3& 0.41& 21.29& GMOS\_N\_r& 0.3312& 0.0029& 0.0000& 0.0000& 0.0000\\ 
& 29.4996 & $65.7203$& 13.7& 0.39& 19.83& GMOS\_N\_r& 0.0656& 0.0123& 0.0000& 0.0000& 0.0000\\ 
& 29.5129 & $65.7192$& 17.1& 0.37& 20.03& GMOS\_N\_r& 0.1202& 0.0101& 0.0000& 0.0000& 0.0000\\ 
& 29.5136 & $65.7181$& 16.3& 0.42& 19.42& GMOS\_N\_r& 0.0598& 0.0190& 0.0000& 0.0000& 0.0000\\ 
& 29.5052 & $65.7125$& 15.7& 0.43& 18.15& GMOS\_N\_r& 0.0141& 0.0763& 0.0000& 0.0000& 0.0000\\ 
& 29.5122 & $65.7171$& 13.5& 0.47& 19.02& GMOS\_N\_r& 0.0274& 0.0291& 0.0000& 0.0000& 0.0000\\ 
& 29.5015 & $65.7124$& 15.7& 0.16& 21.13& GMOS\_N\_r& 0.2748& 0.0034& 0.0000& 0.0000& 0.0000\\ 
& 29.4933 & $65.7138$& 18.0& 0.33& 20.19& GMOS\_N\_r& 0.1541& 0.0086& 0.0000& 0.0000& 0.0000\\ 
& 29.5110 & $65.7136$& 16.4& 0.19& 22.17& GMOS\_N\_r& 0.6012& 0.0013& 0.0000& 0.0000& 0.0000\\ 
& 29.5021 & $65.7130$& 13.6& 0.59& 21.73& GMOS\_N\_r& 0.3480& 0.0019& 0.0000& 0.0000& 0.0000\\ 
& 29.5029 & $65.7218$& 18.3& 0.28& 21.64& GMOS\_N\_r& 0.5053& 0.0021& 0.0000& 0.0000& 0.0000\\ 
FRB180924& 326.1054 & $-40.9002$& 0.8& 1.31& 21.32& VLT\_FORS2\_g& 0.0118& 0.8723& 0.9889& 0.0000& 0.0000\\ 
& 326.1042 & $-40.9002$& 2.9& 0.81& 24.27& VLT\_FORS2\_g& 0.1973& 0.0683& 0.0111& 0.0000& 0.0000\\ 
& 326.1062 & $-40.8993$& 3.8& 0.50& 25.47& VLT\_FORS2\_g& 0.5390& 0.0278& 0.0000& 0.0000& 0.0000\\ 
& 326.1017 & $-40.8998$& 9.5& 0.46& 25.30& VLT\_FORS2\_g& 0.9807& 0.0316& 0.0000& 0.0000& 0.0000\\ 
FRB181112& 327.3486 & $-52.9709$& 0.4& 0.67& 21.49& VLT\_FORS2\_I& 0.0227& 0.0784& 0.8300& 0.0000& 0.0000\\ 
& 327.3496 & $-52.9696$& 5.4& 1.06& 19.10& VLT\_FORS2\_I& 0.0073& 0.8646& 0.1700& 0.0000& 0.0000\\ 
& 327.3484 & $-52.9729$& 7.0& 0.58& 22.01& VLT\_FORS2\_I& 0.1646& 0.0484& 0.0000& 0.0000& 0.0000\\ 
& 327.3467 & $-52.9727$& 7.4& 0.32& 24.05& VLT\_FORS2\_I& 0.6612& 0.0086& 0.0000& 0.0000& 0.0000\\ 
FRB190102& 322.4149 & $-79.4756$& 0.5& 0.86& 20.73& VLT\_FORS2\_I& 0.0038& 0.8425& 1.0000& 0.0000& 0.0000\\ 
& 322.4173 & $-79.4773$& 5.9& 0.55& 22.54& VLT\_FORS2\_I& 0.1623& 0.1575& 0.0000& 0.0000& 0.0000\\ 
FRB190523& 207.0642 & $72.4706$& 3.4& 0.71& 22.13& LRIS\_R& 0.1158& 0.1974& 0.8153& 0.0000& 0.0000\\ 
& 207.0654 & $72.4681$& 5.8& 0.61& 22.82& LRIS\_R& 0.2986& 0.1070& 0.1777& 0.0000& 0.0000\\ 
& 207.0589 & $72.4691$& 6.9& 0.72& 20.78& LRIS\_R& 0.0664& 0.6956& 0.0070& 0.0000& 0.0000\\ 
FRB190608& 334.0203 & $-7.8988$& 2.5& 1.66& 17.60& VLT\_FORS2\_I& 0.0005& 0.9930& 1.0000& 0.0000& 0.0000\\ 
& 334.0185 & $-7.8986$& 5.0& 0.30& 24.83& VLT\_FORS2\_I& 0.5373& 0.0010& 0.0000& 0.0000& 0.0000\\ 
& 334.0186 & $-7.8969$& 6.6& 0.26& 25.28& VLT\_FORS2\_I& 0.8461& 0.0007& 0.0000& 0.0000& 0.0000\\ 
& 334.0187 & $-7.8959$& 9.5& 0.45& 22.76& VLT\_FORS2\_I& 0.4081& 0.0053& 0.0000& 0.0000& 0.0000\\ 
FRB190611& 320.7429 & $-79.3973$& 2.0& 0.50& 22.35& GMOS\_S\_i& 0.0267& 0.3324& 0.9990& 0.0000& 0.0000\\ 
& 320.7495 & $-79.3972$& 3.1& 0.27& 25.87& GMOS\_S\_i& 0.5297& 0.0206& 0.0006& 0.0000& 0.0000\\ 
& 320.7439 & $-79.3985$& 3.3& 0.26& 24.91& GMOS\_S\_i& 0.3446& 0.0412& 0.0004& 0.0000& 0.0000\\ 
& 320.7539 & $-79.3979$& 5.7& 0.65& 23.63& GMOS\_S\_i& 0.3518& 0.1116& 0.0001& 0.0000& 0.0000\\ 
& 320.7383 & $-79.3977$& 4.8& 0.36& 23.44& GMOS\_S\_i& 0.2267& 0.1304& 0.0000& 0.0000& 0.0000\\ 
& 320.7346 & $-79.3988$& 8.4& 0.53& 23.36& GMOS\_S\_i& 0.5033& 0.1392& 0.0000& 0.0000& 0.0000\\ 
& 320.7541 & $-79.3965$& 7.0& 0.18& 26.52& GMOS\_S\_i& 0.9936& 0.0133& 0.0000& 0.0000& 0.0000\\ 
& 320.7569 & $-79.3991$& 9.4& 0.52& 23.65& GMOS\_S\_i& 0.6643& 0.1102& 0.0000& 0.0000& 0.0000\\ 
& 320.7364 & $-79.3998$& 9.9& 0.41& 24.56& GMOS\_S\_i& 0.9166& 0.0536& 0.0000& 0.0000& 0.0000\\ 
& 320.7319 & $-79.3970$& 9.2& 0.27& 25.04& GMOS\_S\_i& 0.9551& 0.0373& 0.0000& 0.0000& 0.0000\\ 
& 320.7587 & $-79.3986$& 9.5& 0.17& 26.91& GMOS\_S\_i& 1.0000& 0.0103& 0.0000& 0.0000& 0.0000\\ 
FRB190614& 65.0743 & $73.7068$& 1.3& 0.41& 24.01& LRIS\_I& 0.0552& 0.1944& 0.5825& 0.0000& 0.0000\\ 
& 65.0738 & $73.7064$& 2.2& 0.40& 22.79& LRIS\_I& 0.0386& 0.5335& 0.4175& 0.0000& 0.0000\\ 
& 65.0705 & $73.7075$& 5.7& 0.33& 24.26& LRIS\_I& 0.4949& 0.1593& 0.0000& 0.0000& 0.0000\\ 
& 65.0817 & $73.7079$& 7.5& 0.17& 26.35& LRIS\_I& 0.9945& 0.0350& 0.0000& 0.0000& 0.0000\\ 
& 65.0691 & $73.7081$& 8.2& 0.28& 25.21& LRIS\_I& 0.9387& 0.0779& 0.0000& 0.0000& 0.0000\\ 
FRB190711& 329.4194 & $-80.3581$& 0.5& 0.46& 22.93& GMOS\_S\_i& 0.0108& 0.4782& 0.9995& 0.0000& 0.0000\\ 
& 329.4187 & $-80.3586$& 2.1& 0.26& 24.88& GMOS\_S\_i& 0.1471& 0.1010& 0.0005& 0.0000& 0.0000\\ 
& 329.4143 & $-80.3570$& 4.7& 0.22& 24.69& GMOS\_S\_i& 0.4654& 0.1163& 0.0000& 0.0000& 0.0000\\ 
& 329.4117 & $-80.3571$& 5.7& 0.29& 24.88& GMOS\_S\_i& 0.6545& 0.1010& 0.0000& 0.0000& 0.0000\\ 
& 329.4190 & $-80.3595$& 5.3& 0.25& 23.97& GMOS\_S\_i& 0.3682& 0.2036& 0.0000& 0.0000& 0.0000\\ 
FRB190714& 183.9795 & $-13.0212$& 1.0& 0.95& 19.48& VLT\_FORS2\_I& 0.0012& 0.7998& 1.0000& 0.0000& 0.0000\\ 
& 183.9797 & $-13.0193$& 6.1& 0.54& 23.71& VLT\_FORS2\_I& 0.3925& 0.0156& 0.0000& 0.0000& 0.0000\\ 
& 183.9787 & $-13.0229$& 7.4& 0.60& 21.22& VLT\_FORS2\_I& 0.0772& 0.1393& 0.0000& 0.0000& 0.0000\\ 
& 183.9797 & $-13.0230$& 7.0& 0.31& 24.36& VLT\_FORS2\_I& 0.6494& 0.0093& 0.0000& 0.0000& 0.0000\\ 
& 183.9795 & $-13.0234$& 8.7& 0.50& 22.71& VLT\_FORS2\_I& 0.3425& 0.0361& 0.0000& 0.0000& 0.0000\\ 
FRB191001& 323.3525 & $-54.7487$& 3.9& 1.36& 17.82& VLT\_FORS2\_I& 0.0009& 0.5074& 0.7174& 0.0000& 0.0000\\ 
& 323.3492 & $-54.7483$& 5.3& 1.47& 17.85& VLT\_FORS2\_I& 0.0015& 0.4920& 0.2826& 0.0000& 0.0000\\ 
& 323.3501 & $-54.7496$& 7.3& 0.27& 25.11& VLT\_FORS2\_I& 0.8690& 0.0005& 0.0000& 0.0000& 0.0000\\ 
FRB200430& 229.7064 & $12.3766$& 0.9& 0.72& 21.19& LRIS\_I& 0.0056& 0.9566& 1.0000& 0.0000& 0.0000\\ 
& 229.7088 & $12.3778$& 8.9& 0.38& 24.82& LRIS\_I& 0.9123& 0.0434& 0.0000& 0.0000& 0.0000\\ 
\hline 
\enddata 
\end{deluxetable*}

\clearpage
\section{Future Directions and Analyses}
\label{sec:future}

This paper and the accompanying code base\footnote{https://github.com/FRBs/astropath}
provide a new methodology to make probabilistic associations of transients
to hosts (PATH).  While we were motivated by FRB science, the general
framework is agnostic to transient type.  Therefore, we anticipate
it will be applied to GRBs, GW events, Type~Ia SNe, and many other
transients.  We stress further that because it is fully probabilistic,
the outputs may be coupled to other likelihood frameworks developed
to constrain, e.g.\ progenitor models or cosmology.

Applied to 13 well-localised FRBs, our results identify nine secure
host galaxies, with posterior probabilities $>0.95$ of being the true host.
We have shown using a suite of `sandbox' simulations that this identification is reliable under a wide range of true FRB host galaxy distributions. This allows a reliable data set to be used when analysing host galaxy properties, or using FRBs for cosmology. Furthermore, by assigning a quantitative probability to individual hosts, we allow even non-secure hosts associations to be used for statistical purposes, with appropriate weighting.

Using these data-sets, we tentatively identify relations between FRB DMs, and host galaxy redshifts, magnitudes, and luminosities. Our results disfavour FRBs as having large offsets from their host galaxies, and we exclude more than one FRB considered as having an unseen host ($p \lesssim 0.01$). Thus we can conclusively answer the oft-asked question ``could the true host galaxies be missed?'' with a ``no''.

Regarding FRBs, future work will include:
 (i) leveraging the next set of $\sim 10$ FRBs to further refine the
 priors and offset function; 
 and 
 (ii) expanding the formalism to include additional observables
 (see the Appendix).
The latter may require obtaining additional data or performing
additional analyses (e.g.\ photo-$z$ estimates) than the 
simple flux and angular sizes considered here.
One may also introduce and test priors motivated by
progenitor models.
Last, the analysis can inform observing strategies to optimize
the probability of a secure association as a function of anticipated
FRB redshift, localization error, and imaging quality.


\section*{Acknowledgements}
We thank C. Kilpatrick and J. Bloom for helpful
discussions.
The Fast and Fortunate for FRB
Follow-up team acknowledges support from 
NSF grants AST-1911140 and AST-1910471. 
A.T.D. is the recipient of an ARC Future Fellowship (FT150100415). K.A. acknowledge support from NSF grant AAG-1714897.
TB gratefully acknowledges support from NSF via grants AST-1909709 and AST-1814778.
C.W.J. acknowledges support by the Australian Government through the Australian Research Council's Discovery Projects funding scheme (project DP210102103).
We thank S. Ryder and L. Marnoch for sharing the reduced
VLT/FORS2 image around FRB~190608 in advance of publication.
This work is partly based on observations collected at the European
Southern Observatory 
under ESO programmes 0102.A-0450(A), 0103.A-0101(A), 0103.A-0101(B) and 105.204W.001.





\end{document}